\documentclass[a4paper,prd,twocolumn,showpacs,superscriptaddress,floatfix,nofootinbib]{revtex4-2}

\usepackage{preamble}
\usepackage{orcidlink}
\usepackage{lipsum}
\usepackage{epsf}
\usepackage{epsfig}
\usepackage{amssymb,amsmath,amsfonts} 
\usepackage{amssymb}
\usepackage{times}
\usepackage{pdfpages}
\makeatletter
\AtBeginDocument{\let\LS@rot\@undefined}
\makeatother

\usepackage{hyperref} \hypersetup{
  colorlinks=true,        
  linkcolor=blue,         
  citecolor=cyan,         
}

\usepackage{graphicx}  
\usepackage{dcolumn}   
\usepackage{bm}        
\usepackage{amsfonts,amsmath,amssymb,mathrsfs}
\usepackage{color}
\usepackage{natbib,times}
\usepackage{mathtools}
\usepackage{xspace}

\newcommand{\fuka}{\texttt{FUKA}\xspace}

\newcommand{\cf}{cf.,~}
\newcommand{\ie}{i.e.,~}
\newcommand{\eg}{e.g.,~}

\begin{document}
\title{Black hole--neutron star binaries with high spins and large mass asymmetries:\\
III. Properties of the ejected material and its electromagnetic signatures}
\author{Konrad Topolski\, \orcidlink{0000-0001-9972-7143}}
\affiliation{Hamburg Observatory, Department of Physics, University of
Hamburg, Gojenbergsweg 112, 21029 Hamburg, Germany}
\affiliation{Institut f\"ur Theoretische Physik, Goethe Universit\"at,
Max-von-Laue-Str. 1, 60438 Frankfurt am Main, Germany}

\author{Samuel D. Tootle\, \orcidlink{0000-0001-9781-0496}}
\affiliation{Department of Physics, University of Idaho, Moscow, ID 83843, USA}
\affiliation{Institut f\"ur Theoretische Physik, Goethe Universit\"at,
Max-von-Laue-Str. 1, 60438 Frankfurt am Main, Germany}

\author{Paramvir Singh\, \orcidlink{0000-0003-1006-6970}}
\affiliation{Institut f\"ur Theoretische Physik, Goethe Universit\"at,
Max-von-Laue-Str. 1, 60438 Frankfurt am Main, Germany}
\affiliation{INAF-Osservatorio di Astrofisica e Scienza dello Spazio, Via
  Piero Gobetti 93/3, 40129 Bologna, Italy}

\author{Mattia Bulla\, \orcidlink{0000-0002-8255-5127}}
\affiliation{Department of Physics and Earth Science, University of Ferrara, Via
Saragat 1, I-44122 Ferrara, Italy}
\affiliation{INFN, Sezione di Ferrara, Via Saragat 1, I-44122 Ferrara,
Italy}
\affiliation{INAF - Osservatorio Astronomico d'Abruzzo, Via Mentore Maggini
snc, I-64100 Teramo, Italy.}

\author{Luciano Rezzolla\, \orcidlink{0000-0002-1330-7103}}
\affiliation{Institut f\"ur Theoretische Physik, Goethe Universit\"at,
  Max-von-Laue-Str. 1, 60438 Frankfurt am Main, Germany}
\affiliation{Department of Mathematics, New Uzbekistan University, Tashkent
  100007, Uzbekistan}
\affiliation{School of Mathematics, Trinity College, Dublin 2, Ireland}

\date{\today}

\begin{abstract}
Compact binary systems consisting of a black hole (BH) and a neutron star
(NS) are expected to provide detectable signals both in the gravitational
wave and in the electromagnetic channels, thus representing excellent
sources for multi-messenger astronomy. In this third paper in a series,
we report on the geometric and thermodynamic properties of the purely
dynamical ejecta component modeled in our general-relativistic
magnetohydrodynamical simulations of BHNS systems over a comprehensive
binary parameter space consisting of large mass ratio and high BH
spin. The tidal disruption and dynamical mass ejection is studied by
monitoring the transfer of energy and angular momentum to matter for the
particular case of a highly unequal-mass binary with large BH spin,
extending known results for low-mass-ratio, irrotational BHNS
binaries. By varying the binary parameters, we investigate their
influence on the geometric, thermodynamic, and kinematic properties of
the ejected matter. Furthermore, we employ the nuclear-reaction network
\texttt{SkyNet} to obtain the elemental abundances resulting from the
$r$-process active in the ejecta and study how they depend on the
properties of the progenitor objects. Finally, using the
radiative-transfer code \texttt{POSSIS}, we compute the kilonova
light-curves from the results of our simulations, discuss their
bolometric and optical/near-infrared evolution, and compare them with the
AT2017gfo data and upper-limits for S190814bv, finding overall
consistency.
\end{abstract}
\maketitle

\section{Introduction}
\label{sec:introduction}

The detection of the gravitational-wave (GW) event
GW170817~\cite{Abbott2017, Abbott2017b} and its electromagnetic (EM)
counterparts -- the short gamma-ray burst (sGRB)
GRB170817A~\cite{Goldstein2017, Savchenko2017} and the kilonova (KN)
AT2017gfo~\cite{Coulter2017, Chornock2017, Nicholl2017,
  Cowperthwaite2017, Pian2017, Smartt2017, Tanvir2017, Tanaka2017,
  Valenti2017} delivered an unprecedented amount of information about the
final stages of a binary neutron star (BNS) coalescence, allowing to put
new constraints on the neutron star (NS) mass-radius relation and on the
equation of state (EOS) of dense nuclear matter~\cite{Margalit2017,
  Rezzolla2017, Ruiz2017, Most2018, Burgio2018b, De2018, Landry2018,
  Baiotti:2019sew, Raaijmakers2019, Shibata2019, Essick2021, Annala:2022,
  Burgio2021, Capano2020, Chatziioannou2020a, Raaijmaers2021b,
  Lattimer2021a, Breschi2021, Nathanail2021}. Moreover, it established
the post-merger configuration as a viable jet engine and sGRB
progenitor~\cite{Goldstein2017, Savchenko2017, Abbott2017b, Lazzati2017c,
  Mooley2018b}, as well as identified~\cite{Pian2017, Smartt2017} the
ejected material as an unambiguous location of $r$-process
nucleosynthesis~\cite{Tanaka2017, Valenti2017}, enabling direct
confirmation of the presence of heavy elements such as
Sr~\cite{Watson2019}.

A related class of compact binary mergers expected to produce
multi-messenger signals consists of a black hole (BH) and a NS, which
have thus far been represented by two confirmed GW detections by the
LIGO-Virgo-Kagra (LVK) collaboration~\cite{Acernese:2014,Aasi:2014} --
GW200105 and GW200115~\cite{Abbott2021} -- and several further BHNS
candidates identified during subsequent observing campaigns, up until
the recently completed O4 run~\cite{LIGOScientific2025,
  LIGOScientific2026}, with a notable mention of the hitherto unique ``mass-gap'' NSBH system GW230529~\cite{LIGOScientific2024}.
  Extensive EM follow-ups were performed for these
events~\cite{Dichiara2021, LIGOScientific2023, Ronchini2024} (gamma-ray searches,
wide-field optical searches, as well as X-ray/radio observations and
neutrino limits), producing several upper limits on prompt GRB and KN
emission. However, no secure EM counterpart to a BHNS merger has been
confirmed yet. Implications of a BHNS multi-messenger detection include,
but are not limited to, constraining further the EOS of dense nuclear
matter~\cite{Lackey2013, Mathias2023,Raaijmakers2023}, verifying the
theoretical predictions of relativistic jet launching and matter ejection
in BHNS mergers~\cite{Rezzolla:2010, Deaton2013, Paschalidis2014,
  Foucart2015a, Foucart2017b, Kyutoku2018, Brege2018, Ruiz2018,
  Most2021a, Hayashi2022, Gottlieb2023a, Hayashi2023}, as well as
evaluating the relative contribution of the channel involving $r$-process
nucleosynthesis in the BHNS ejecta to the observed solar abundances of
heavy elements~\cite{Wanajo2022}.

KNe from BHNS binaries are multi-component transients, associated with
various ejection timescales, similarly to the picture valid for BNS
mergers~\cite{Korobkin2012, Metzger2019a, Perego2017, Radice2018,
  Cowperthwaite2017, Kasen2017, Tanvir2017, Tanaka2017, Breschi2021}.
Numerical studies over the last decade~\cite{Fernandez2015b,
  Kawaguchi2016, Fernandez2017, Li2019d, Barbieri2019a, Andreoni2020,
  Anand2021, Darbha2021, Markin2023, Mathias2023, Kunnumkai2024,
  Kawaguchi2024a, Markin2026} found that BHNS mergers are typically dim
in the optical band, but bright in the infrared spectrum, as a direct
cause of the abundant production of lanthanides~\cite{Kasen2017} during
the $r$-process, leading to effects such as photon reprocessing in the
ejecta. In particular, disruptive BHNS mergers for binaries of
intermediate mass ratio and considerable BH spins~\cite{Kawaguchi2024a},
or those where the primary BH is a mass-gap object~\cite{Kunnumkai2024}
may be accompanied by KNe which are fainter, but likely more enduring
than the one observed in AT2017gfo.

Existing literature on the $r$-process nucleosynthesis and KNe in
BHNS and BNS mergers frequently assumes an idealized ejecta distribution
and composition, derived from analytical models and informed by results
of numerical-relativity simulations, and subsequently draws broad
conclusions from large data ensembles obtained by varying the different
unbound mass components and their properties (see, \eg \cite{Barnes2013,
  Kawaguchi2016, Bulla2019, Anand2021,Koehn2025, Hu2025}). Here, we take
a different, \textit{ab-initio} approach, combined with a case-by-case
analysis, tying together the geometric, kinematic and thermodynamic
features of dynamical ejecta in our simulations -- and hence the
$r$-process yields and the KN observables -- with the underlying binary
parameters. Namely, in this paper, which is the third in a
series~\cite{Topolski2024b, Topolski2024c}, we investigate the mechanism
of dynamical mass ejection in case of BHNS simulations carried out in
Paper II~\cite{Topolski2024c}, focusing on the BH spin- and mass-ratio
dependence of the ejecta mass, its geometrical properties -- such as the
angular distribution and collimation -- the kinematic and thermodynamic
properties, and finally the results of nuclear-reaction networks and
radiative-transfer simulations that lead to EM-related observables.

Below we briefly describe the structure of the paper. In
Sec.~\ref{sec:sim_overview} we recall the setup of the simulations
performed in Paper II, and describe the probed space of the binary
parameters. Before addressing the properties of the unbound matter, in
Sec.~\ref{sec:dyn_mass_EJ} we investigate in detail the mechanism behind
the formation of dynamical ejecta, using as an example a particularly
high-mass ratio binary. Following this discussion, we present the bulk
thermodynamic-kinetic properties of the ejecta as a function of binary
mass ratio $Q:=M_{_{\rm BH}}/M_{_{\rm NS}}$ and dimensionless BH spin
$\chi_{_{\rm BH}}$ in Sec.~\ref{sec:bulk_prop}, including a statistical
and spatial characterization of the material in terms of electron
fraction $Y_e$, velocity $v$ and entropy $s$. Therein, we also discuss
our choice of the detector radius, where the ejecta is measured for the
purpose of extracting artificial tracers. In
Sec.~\ref{sec:ejecta_angular_distribution} we investigate the angular
extent of the dynamical ejecta as extracted on spherical detector
surfaces. The $r$-process nucleosynthesis and elemental abundances
obtained through the \texttt{SkyNet} code are presented in
Sec.~\ref{sec:rprocess_nucleosynthesis}, with their variance explained in
a dedicated
Appendix~\ref{subsec:rprocess_nucleosynthesis_entropy_vel_dep}. Finally,
Sec.~\ref{sec:radiative_transfer_kilonova} presents bolometric and
photometric light-curves of the kilonovae in our study obtained using the
radiative-transfer code \texttt{POSSIS}. Therein, apart from discussing
the impact of $Q$ and $\chi_{_{\rm BH}}$ on the luminosity and
time-evolution of computed light-curves, we discuss the dependence on the
viewing angle as well as compatibility with upper limits on the EM
detection from select candidate observations. We close by presenting our
conclusions and outlook in Sec.~\ref{sec:conclusions}. Throughout the
paper, we use geometrical units $G = c = 1$, where $G$ and $c$ are the
gravitational constant and the speed of light, respectively. Greek
(Latin) indices run from 0 (1) to 3.

\section{Simulations' overview}
\label{sec:sim_overview}

We here briefly recall the parameter space covered by the simulations of
Ref.~\cite{Topolski2024c} and the codes used to obtain the dynamical
ejecta. The entirety of the most important information about the
simulations is contained in Tab.~\ref{tab:ID_properties}. In total, we
investigate $6$ BHNS configurations at medium-to-high mass ratio $Q$ and
varying BH spin $\chi_{_{\rm BH}}$, excluding the rather trivial plunge
scenario \texttt{Q4.chi0.0} in ~\cite{Topolski2024c} due to the
negligible amount of matter left after the coalescence. Otherwise, our
parameter space consists of two overlapping sets: binaries at a fixed
mass ratio $Q=4$ and increasing spin $\chi_{_{\rm BH}} = [0.4, 0.6,
  0.8]$, or binaries at a fixed (and quite substantial) BH dimensionless
spin $\chi_{_{\rm BH}}=0.8$ and increasing mass ratio $Q = [4, 5, 6,
  7]$. The ADM mass of the NS in each case is $M_{_{\rm
    NS}}=1.4\,M_{\odot}$, corresponding to baryonic mass $M_{\rm b,
  NS}=1.53\,M_{\odot}$, where the DD2 EOS~\cite{typel:2009sy,
  hempel:2009mc} is used. While this coverage of the parameter space is
not exhaustive, it reliably covers the region of the BHNS parameter space
where the remaining baryon mass after the coalescence is sufficiently
high to produce an EM counterpart. In addition to the aforementioned ID
properties, we also denote each coalescence as a weak or strong tidal
disruption (WTD and STD, respectively). A rigorous classification into
three subtypes relying purely on information from BHNS initial data
sequences was given in Refs.~\cite{Topolski2024b, Topolski2024c}, where,
we recall, we computed the GW frequencies at the onset of mass-shedding
and at the crossing of effective ISCO (innermost stable circular orbit),
$f_{_{\rm MS}}$, and $f_{_{\rm ISCO}}$, respectively. We then introduced
the so-called \textit{quasi-equilibrium survival time} \hbox{$\tau_{\rm
    sur}^{^{\rm QE}} := g_{_{\rm ISCO}}(\chi_{_{\rm BH}}) (1/f_{_{\rm
      MS}} - 1/f_{_{\rm ISCO}})$}, namely, a timescale associated with
the difference in (the inverse of) these two frequencies [here, $g_{_{\rm
      ISCO}}(\chi_{_{\rm BH}})$ is a function that captures the
  dependence of the ISCO on BH spin]. In this way, it was possible to
classify the merger as a ``strong disruption'' if $M_{\rm tot}^{-1} \,
\tau_{\rm sur}^{^{\rm QE}}\gtrsim 20$ and as a ``weak disruption'' if
\hbox{$0 \lesssim M_{\rm tot}^{-1} \, \tau_{\rm sur}^{^{\rm QE}}\lesssim
  20$}, where $M_{\rm tot}$ is the total binary mass measured at infinite
separation. Furthermore, in this effective classification, a ``plunge'',
\ie when the NS falls into the BH without forming an accretion disk,
simply corresponds to the case in which $M_{\rm tot}^{-1} \, \tau_{\rm
  sur}^{^{\rm QE}} \lesssim 0$. Importantly, this classification matched
well the classification scheme of Refs.~\cite{Kyutoku2011,Pannarale2011,
  Pannarale2013a, Pannarale2015} calibrated on dynamical simulations and
captures the impact of possible disruption on the excitation of
quasi-normal modes~\cite{Topolski2024c} already from the properties of
the initial data.

The initial data given in Tab.~\ref{tab:ID_properties} is obtained with
the initial-data solver \fuka \cite{Papenfort2021b, Tootle2024a} based on
the \texttt{Kadath} spectral-solver library~\cite{Grandclement09}. The
constraint equations on an initial hypersurface are cast and solved in
the XCTS formulation, and thus consistent with the quasi-equilibrium
assumption on the initial data~\cite{Pfeiffer:2002iy, York99}. To evolve
the binaries through the inspiral and early post-merger, we used the
infrastructure of \texttt{EinsteinToolkit}~\cite{Loffler:2011ay} with the
fixed-mesh box-in-box refinement framework
\texttt{Carpet}~\cite{Schnetter:2003rb}. The \texttt{Antelope} code
integrated into the~\texttt{EinsteinToolkit}~\cite{Most2019b} evolves the
metric variables by utilizing a 4th-order, up-wind finite differencing
scheme. For our simulations, we opted for the moving-punctures based
CCZ4~\cite{Alic:2011a, Alic2013} or Z4c~\cite{Bernuzzi:2009ex,
  Hilditch2012} formulation with damping coefficients $\kappa_{1}=0.02$
and $\kappa_{2}=0$, or, for simulations with $Q=[6,7]$, the BSSN
formulation~\cite{Nakamura87, Shibata95, Baumgarte99}, which we found
more suitable to ensure adequate BH mass conservation (see also the
comments in~\cite{Topolski2024c}). The evolution equations for the
magnetized fluid are solved using the Frankfurt-IllinoisGRMHD
(\texttt{FIL}) code~\cite{Most2019b}, also implemented as a thorn within
the~\texttt{EinsteinToolkit} suite. While we have endowed the NS
with a magnetic field of poloidal topology, the magnetic field is not
dynamically important in the context of this paper, with its maximum
values equal to $\vert B\vert \approx 10^{13}\,\rm{G}$ before the merger
and decreasing by an order of magnitude in the post-merger phase (see
also~\cite{Topolski2024c}).

Differently from Paper II, we do not consider here the
eccentricity-reduced dataset~\texttt{Q4.chi0.8.er} obtained via the
eccentricity-reducing approach of Ref.~\cite{Papenfort2021b}. The
rationale for this is the absence of measurable differences in terms of
the dynamical ejecta properties between the eccentricity-reduced and the
non-reduced setup. All of the BHNS simulations were performed with three
sets of refinement-level hierarchies, centered at both objects and at the
grid origin, each consisting of seven refinement levels. Both of the
objects are then endowed with an additional, eighth refinement level. The
resulting grid-spacing around the BH and the NS is thus $147\,{\rm
  m}$. The total domain extent is $(3025\, {\rm km})^3$ and reflection
symmetry with respect to the $z=0$ plane is imposed.

\begin{table}[t]
  \begin{ruledtabular}
    \begin{tabular}{l|cccrccc}
      binary & $Q$ & $\chi_{_{\rm BH}}$ & $M_{_{\rm BH}}$ &
      $M_{\rm tot}$ & $d_{0}$ & $e$ & scenario\\
      &&& $[M_{\odot}]$ & $[M_{\odot}]$ & $[M_{\odot}]$ &
      $[10^{-2}]$ & \\
      \hline
      \texttt{Q4.chi0.4} & $4$ & $0.4$ & $5.6$ & $7.0$  & $56$ & $3.2$ & WTD\\
      \texttt{Q4.chi0.6} & $4$ & $0.6$ & $5.6$ & $7.0$  & $56$ & $2.8$ & STD\\
      \texttt{Q4.chi0.8} & $4$ & $0.8$ & $5.6$ & $7.0$  & $56$ & $2.5$ & STD\\
      \texttt{Q5.chi0.8} & $5$ & $0.8$ & $7.0$ & $8.4$  & $56$ & $3.7$ & STD\\
      \texttt{Q6.chi0.8} & $6$ & $0.8$ & $8.4$ & $9.8$  & $60$ & $4.6$ & WTD\\
      \texttt{Q7.chi0.8} & $7$ & $0.8$ & $9.8$ & $11.2$ & $64$ & $5.5$ & WTD\\
    \end{tabular}
  \end{ruledtabular}
  \caption{Properties of the initial data and characterization of the
    disruption strength. From left to right we report: the inverse mass
    ratio $Q:= q^{-1} =M_{_{\rm BH}} / M_{\rm NS}$, the total mass of the
    binary $M_{\rm tot}$ measured at infinite separation, the initial BH
    spin $\chi_{_{\rm BH}}$, the initial separation $d_{0}$, the BH
    Christodoulou mass $M_{_{\rm BH}}$, and the estimated eccentricity
    $e$ based on the dynamical evolution. The outcome of the merger as a
    strong or weak tidal disruption is denoted with STD or WTD,
    respectively.}
  \label{tab:ID_properties}
\end{table}
%

\subsection{Dynamical mass ejection - energy and angular momentum transfer}
\label{sec:dyn_mass_EJ}

In this section we perform the analysis of the transfer of energy and
angular momentum from the binary to the disrupted matter, which allows to
understand the main mechanism behind the creation of dynamical ejecta.
The mechanism of tidal disruption in this case displays similar patterns
as in the $Q<7$ cases, but features a number of interesting differences
that we wish to highlight. The most important difference between the
\texttt{Q7.chi0.8} and simulations at lower mass ratios is the reduced
presence of matter around the BH immediately after the coalescence [\cf
  the results of Paper II~\cite{Topolski2024c}]. There are at least three
reasons behind this behavior. Firstly, the ISCO location for this BH is
now moved significantly outward and close to the orbit where the mass
shedding begins. Secondly, since the high BH mass promotes tidal
disruption -- we recall that in a Newtonian picture, the tidal-disruption
radius is given by $r_{\rm dis}\propto (M_{_{\rm BH}}/M_{_{\rm
    NS}})^{1/3}R_{_{\rm NS}}$ -- any surviving bound material that is
disrupted has to be far enough from the ISCO. Third, the tidal transfer
of energy to the disrupted material follows along a more elongated tidal
tail than in the case of the $Q<7$ binaries, and the tidal tail performs
more rotations around the BH before the tidal disruption is finished, as
marked by the collision of its front and back parts. This facilitates the
transfer of angular momentum and delays the accumulation of the material
in the form of an accretion disk. As a consequence, the formation of a
semi-coherent disk is delayed and can be estimated at around $10\, {\rm
  ms}$ after the time of merger, defined as the time when the amplitude
of the dominant gravitational strain mode reaches its maximum
(see~\cite{Topolski2024c}). About $5\, {\rm ms}$ after the coalescence,
the amount of matter closer to the BH than $\approx 120\, {\rm km}$ is
substantially depleted. The slope of the resulting $M_{\rm disk}(t)$
curve, as noticed in~\cite{Topolski2024c}, implies a prolonged fallback
time, \ie the time needed for $M_{\rm disk}$ to be approximately equal to
$M_{\rm bound}$ (see also~\cite{Musolino2024}).

To perform the analysis of the energy and angular-momentum transfer for
the \texttt{Q7.chi0.8} run, we follow the methodology introduced in
Ref.~\cite{Hayashi2020}, of which we recall below the key
components. Close to the merger time and for considerable mass ratios,
the spacetime increasingly resembles that of an isolated rotating BH with
a perturbation caused by the presence of a secondary -- in this case, the
NS. This implies the existence of two approximate Killing vectors, which
lead to conserved quantities for geodesic motion. The time-like Killing
vector, \ie the coordinate time vector $\boldsymbol{\partial_{t}}$
implies the existence of conserved specific energy $\tilde{E} = -u_{t}:=
(\partial_{t})_{\mu}u^{\mu}$, where $u^{\mu}$ is the four-velocity of the
fluid~\cite{Rezzolla_book:2013}. Similarly, the space-like Killing vector
associated with axisymmetry can be expressed with the help of the
$\boldsymbol{\partial_{\varphi}}$ vector centered at the BH, which leads
to a conserved specific angular momentum $\tilde{J} = u_{\varphi} :=
(\partial_{\varphi})_{\mu}u^{\mu} = x\,u_{y}-y\,u_{x}$, where the
coordinates $x,y$ are again centered on the BH. Of course, in the case of
a binary, neither $\boldsymbol{\partial_{t}}$ nor
$\boldsymbol{\partial_{\varphi}}$ are strict Killing vectors, but they
closely resemble them in the case of a large enough mass ratio and during
the late inspiral. Furthermore, while the four-velocity $\boldsymbol{u}$
is not geodesic throughout the inspiral, the assumption of geodesic
motion is reasonable once the NS disrupts, whereupon the motion of fluid
elements can be treated as independent with good
approximation~\cite{Pannarale2010}.

Under these assumptions, the distribution of matter in the
$(\tilde{E},\tilde{J})$ phase-space is computed with the help of the
following integral
\begin{equation}
	\frac{dM_{\rm b}}{d\tilde{E}d\hat{J}} := \int
  D(\tilde{E}',\hat{J}') \,\sqrt{\gamma} d^{3} x\,,
  \label{eq:dMdEdJ}
\end{equation}
where $D := W\,\rho$ is the conserved density ($\rho$ is the
primitive rest-mass density), $W$ the Lorentz factor, and $\gamma$ the
determinant of the spatial metric
$\gamma_{ij}$~\cite{Rezzolla_book:2013}. The integral~\eqref{eq:dMdEdJ}
is performed only for fluid elements in a given range of specific energy
and angular momentum, namely, $\vert \tilde{E} - \tilde{E}' \vert <
\Delta \tilde{E}/2$ and $\vert \hat{J} - \hat{J}'\vert < \Delta
\hat{J}/2$, where $\hat{J} := {\tilde{J}}/{M^{\rm rem}_{\rm Ch}}$ is the
normalized specific angular momentum. We have sampled the phase-space
using a rather fine spacing of $\Delta\tilde{E} = 0.005$ and $\Delta
\hat{J}=0.025$, which leads to $240$ and $280$ bins in $\tilde{E}\in\{0,
1.2\}$ and $\hat{J}\in \{0,7.0 \}$, respectively.

\begin{figure*}
  \centering
  \includegraphics[width=0.75\textwidth]{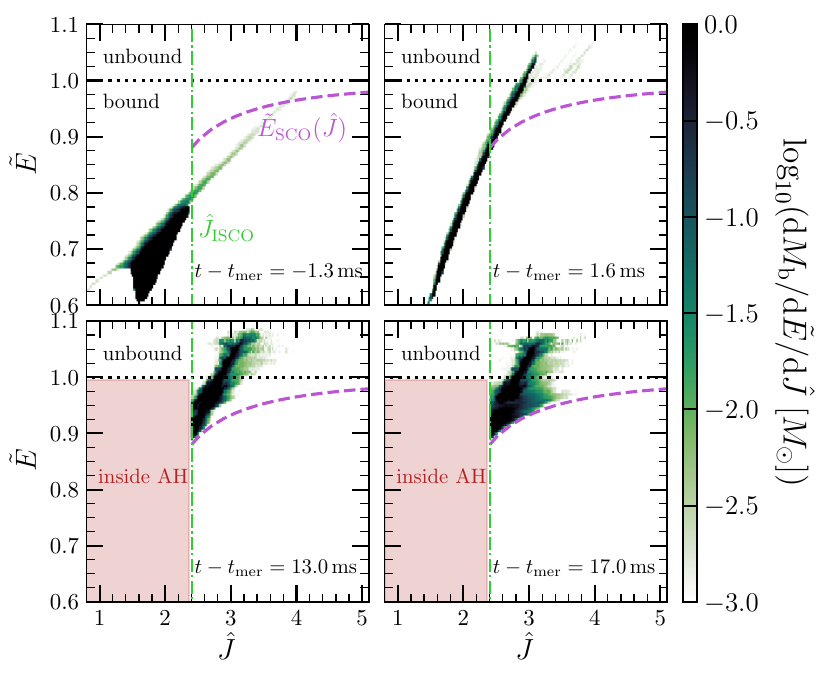}
  \caption{Distribution of the baryonic mass $M_{\rm b}$ (see colorbar)
    in the phase-space spanned by the specific energy $\tilde{E}$ and the
    normalized specific angular momentum $\hat{J} = \tilde{J} / M^{\rm
      mer}_{\rm Ch}$, for the \texttt{Q7.chi0.8} binary. The red-shaded
    area represents a part of the phase space occupied by bound,
    low-angular momentum stellar material which is either within the
    apparent horizon already, or will inevitably enter it. }
  \label{fig:dM_overdEdJ_q7}
\end{figure*}

The computed rest-mass distribution in the phase-space of the specific
energy $\tilde{E}$ and of the normalized specific angular momentum
$\hat{J}$ is presented in Fig.~\ref{fig:dM_overdEdJ_q7} at times
$t-t_{\rm mer}=(-1.3,1.6,13.0,17.0)\, {\rm ms}$. The phase-space is
divided horizontally into regions of bound $\tilde{E}<1$ and unbound
$\tilde{E}>1$ matter, as well as vertically into regions with
(normalized) specific angular momentum higher or lower than that of the
ISCO $\hat{J}_{\rm ISCO}$, shown as a vertical dash-dot green line. The
purple dashed curve $E_{\rm SCO}(\hat{J})$ represents the relation
between $\tilde{E}$ and $\hat{J}$ for stable circular orbits at radii
$r>r_{\rm ISCO}$. The analytical dependence is computed considering the
mass and spin of the BH at the end of the simulation.

Starting from the top left, the first panel in
Fig.~\ref{fig:dM_overdEdJ_q7} refers to the late inspiral, so that the
matter is both bound and possesses angular momentum smaller than
$\hat{J}_{\rm ISCO}$. In the second panel, the NS becomes disrupted and
the tidal tail spreads out across a range of energies and angular
momenta, stretched through a thin region of the phase-space. Part of the
matter in the outer regions of the tidal tail gains energies which
satisfy $\tilde{E}>1$ and thus becomes unbound, while some of it is
promptly accreted. At the time corresponding to the third panel, most of
the promptly accreted matter has been removed from the computational
domain. The matter with $\tilde{E}<1$, $\hat{J}<\hat{J}_{\rm ISCO}$ is
still present in the phase-space but is not reported in the figure
because it will inevitably meet the singularity in the BH interior;
because of this, the corresponding region in the phase-space is shown with
a red-shaded area. During the transition from the third to the fourth
panel that lasts $\approx 4\, {\rm ms}$, bound matter with enough angular
momentum starts settling at the stable circular orbits along the $J_{\rm
  SCO}$ line, while the unbound matter does not change its energy nor
angular momentum substantially as it evolves with an almost ballistic
motion.

What is important to remark here, and is different from the lower
mass-ratio simulations of~\cite{Hayashi2020}, is that a large part of the
bound matter component displayed in the phase-space has not yet settled
along the curve representing stable circular orbits. At the same time,
much of the bound matter below the $\hat{J}_{\rm ISCO}$ is still present,
although it will soon be accreted by the BH. The prominent presence of
matter in this part of the phase-space is the result of the large remnant
BH mass $M_{_{\rm BH}}^{\rm rem} \approx 10.890\,M_{\odot}$, which allows
matter to be in free-fall below the ISCO radius. The differences between
our Fig.~\ref{fig:dM_overdEdJ_q7} and the phase-space representation in
Ref.~\cite{Hayashi2020} -- in which irrotational and lower-mass BHs were
considered -- are therefore the clear imprint on the matter distribution
resulting from the large mass ratio and high BH spins considered here. As
a result, Fig.~\ref{fig:dM_overdEdJ_q7} complements and extends the
analysis performed in~\cite{Hayashi2020}, offering compelling evidence
that large $Q$ in conjunction with significant $\chi_{_{\rm BH}}$ leads
to a delayed formation of an accretion disc.

\section{Results}
\label{sec:results}

\subsection{Bulk properties of the ejecta}
\label{sec:bulk_prop}

We next discuss the bulk properties of the matter ejected for all of our
simulations. To identify the \textit{dynamical ejecta}, we consider the
component of rest-mass expelled from the system in the immediate
aftermath of a BHNS coalescence which satisfies a given
\textit{``unboundness''} criterion, which in our case corresponds to the
geodesic criterion $u_{t} < -1$, where the covariant time component of
the four-velocity $u_{\mu}$ of the fluid is used, with its negative
equivalent to the measured energy at infinity, as well as the more
permissive Bernoulli criterion $hu_{t} < -h_{\min}$, with $h$ the
specific enthalpy (we here take $h_{\min}=-1$). Since $h \geq 1$, it is
clear that the Bernoulli criterion will be in general less restrictive
than the geodesic one, yielding an ejected mass of the material that is
up to a factor of two larger~\cite{Bovard2017}; more elaborate
unboundness criteria can also be derived for axisymmetric spacetimes with
magnetic fields present (see, \eg \cite{Bekenstein1978, Gourgoulhon2011,
  Musolino2024b}).

\begin{table*}[t]
  \renewcommand{\arraystretch}{1.25} 
  \begin{ruledtabular}
    \begin{tabular}{c|crccccccc}
      binary & $\chi_{_{\rm BH}}^{\rm rem} $ & $M_{_{\rm BH}}^{\rm rem}$ &
	    $M^{\rm rem}_{\rm irr}$ & $M_{\rm disk}$ & $M_{\rm ej}$ & $M_{\rm ej}^{\rm geo}$ & $M_{\rm
	b, rem}$ & $\hat{M}_{\rm b, rem}$ & $M_{\rm ej}/M_{\rm b, NS}$  \\
      & & $[M_{\odot}]$ & $[M_{\odot}]$ & $[M_{\odot}]$ & $[10^{-1}
	    M_{\odot}]$& $[10^{-1}M_{\odot}]$ & $[M_{\odot}]$ & $[M_{\rm b, NS}]$ &  $\times 100$ 
      \\
      \hline
	    \texttt{Q4.chi0.4~~~} & $0.679$ &  $6.836$ & $6.365$ & $0.004$     & $0.063$ & $ 0.063 $   & $ 0.024$    & $0.016$      & $0.412$  \\
	    \texttt{Q4.chi0.6~~~} & $0.772$ &  $6.658$ & $6.021$ & $0.114$     & $0.285$ & $ 0.195 $   & $ 0.202$    & $0.132$      & $1.863$ \\
	    \texttt{Q4.chi0.8~~~} & $0.868$ &  $6.582$ & $5.694$ & $0.189$     & $0.580$ & $ 0.427 $   & $ 0.293$    & $0.192$      & $3.790$ \\
	    \texttt{Q5.chi0.8~~~} & $0.861$ &  $7.979$ & $6.930$ & $0.160$     & $0.557$ & $ 0.414 $   & $ 0.275$    & $0.180$      & $3.641$ \\
	    \texttt{Q6.chi0.8~~~} & $0.858$ &  $9.414$ & $8.190$ & $0.103$     & $0.376$ & $ 0.282 $   & $ 0.212$    & $0.138$      & $2.457$ \\
	    \texttt{Q7.chi0.8~~~} & $0.858$ & $10.890$ & $9.474$ & $0.025$     & $0.314$ & $ 0.233 $   & $ 0.108$    & $0.070$      & $2.052$ \\
    \end{tabular}
  \end{ruledtabular}
  \caption{Properties of ejected material as measured at the end of the
    simulations. From left to right we report: the remnant BH spin
    $\chi_{_{\rm BH}}^{\rm rem}$, the corresponding gravitational mass
    $M_{_{\rm BH}}^{\rm rem}$ and irreducible mass $M_{\rm irr}^{\rm
      rem}$, the accretion-disk mass $M_{\rm disk}$, the ejected mass
    according to the Bernoulli $M_{\rm ej}$ and geodesic criterion
    $M_{\rm ej}^{\rm geo}$, the remnant baryonic mass $M_{\rm b, rem}$,
    and its value normalized by the NS initial baryonic mass
    $\hat{M}_{\rm b, rem}$, and, finally, the ratio of the dynamical
    ejecta mass to the initial baryonic mass of the NS $M_{\rm ej}/M_{\rm
      b, NS}$.}
	\label{tab:sim_diag}
\end{table*}

The total ejected rest-mass is defined as usual through an integral
computed outside of the BH apparent horizon
\begin{equation}
  M_{\rm ej} := \int_{h u_{t}<-1} W \rho \sqrt{\gamma} d^{3}x\,.
\end{equation}
Since the integral is limited to matter with $h u_t < -1$, we will ignore
here the material that is ejected with large velocity but is
gravitationally bound. As discussed in Ref.~\cite{Musolino2024}, this
matter will follow a precise mass accretion rate and could be useful to
explain long-term EM emissions from BHNS mergers. Furthermore, we note
that alternative and potentially more precise measurements could be
achieved by introducing a sufficiently large number of tracer particles
as done in Ref.~\cite{Bovard2016}, although the largest uncertainty here
rests with the actual criterion for unboundness chosen.

In Tab.~\ref{tab:sim_diag}, we present the measurements of most important
parameters of the system in the aftermath of the coalescence.
This Table
overlaps in terms of content with the one in~\cite{Topolski2024c}, and we
repeat it for easy overview and reference later on.

Clearly, stronger tidal disruption present in the \texttt{Q4.chi0.8} and
\texttt{Q5.chi0.8} cases promotes the ejection of a substantial amount of
material $\mathcal{O}(0.04\,M_{\odot})$\footnote{Measured according to
the geodesic criterion.}, which is halved when the mass ratio of the
system increases to $Q=7$. The simulations \texttt{Q4.chi0.8} and
\texttt{Q5.chi0.8} feature an amount of dynamical ejecta which differs by
$4\%$, putting these datasets at a plateau maximum of a putative $M_{\rm
  ej}(Q, \chi_{_{\rm BH}}, {\rm EOS})$ function for a fixed BH spin and
EOS, similarly to Fig.~3 in Ref.~\cite{Hayashi2020}. We note that in
principle, larger dynamical ejecta masses are also possible, especially
for NSs of low stellar compactness; indeed at sufficiently high initial
BH spin, mass ratios $Q\gtrsim 6$ may produce unbound material several
times as large as the one we are reporting. For example,
Ref.~\citet{Foucart2014} reports a number of configurations with $M_{\rm
  ej} \gtrsim 0.1 \,M_{\odot}$, including a configuration where $Q\approx
6$, $\chi_{_{\rm BH}}=0.9$ and $\mathcal{C}_{\rm NS}\approx 0.14$, which
produces $M_{\rm ej} = 0.16\,M_{\odot}$, as well as similar
configurations with large $Q\approx 8.3$ where $M_{\rm ej} \approx
0.10-0.15\,M_{\odot}$ albeit the latter were plagued by accuracy losses
due to insufficient resolution in the region away from the BH and thus
most important for modelling the mass ejection mechanism. Similarly,
Ref.~\citet{Kyutoku2015} also reports results with $M_{\rm
  ej}>0.05\,M_{\odot}$ provided that the BH spin is $\chi_{_{\rm
    BH}}\gtrsim 0.5$ and/or the stellar compactness of the NS model is
$\mathcal{C}_{\rm NS} \approx 0.14-0.15$.

The maximum of $M_{\rm ej}(Q)$ for our datasets appears for slightly
higher values of $Q\approx 4-5$ since the BHs in our binaries are
initially spinning with a significant $\chi_{_{\rm BH}}=0.8$, which
favors tidal disruption\footnote{The BHs considered in
Ref.~\cite{Hayashi2020} were irrotational.}. Similarly
to~\cite{Kyutoku2015}, we find that weak disruption does not decrease the
amount of dynamical ejecta at high mass ratios to a degree comparable
with the decrease of the total remnant rest-mass. In other words, the
mass ratio dependence of the total remnant mass $M_{\rm rem}$ features a
more pronounced decline at high mass asymmetries (see the entries of
Tab.~\ref{tab:sim_diag}). Overall, the growth of the ratio $M_{\rm
  ej}/M_{\rm rem}$ with $Q$ is to be expected for weakly disruptive
mergers, since in highly asymmetric BHNS systems the material that is
disrupted and not accreted promptly is necessarily found at large radii
and hence beyond $r_{\rm ISCO}$.

\begin{table*}
  \renewcommand{\arraystretch}{1.25} 
  \begin{ruledtabular}
    \begin{tabular}{c|c|ccc|cccc|ccc|c}
      \hline
      binary & $M_{\text{ej}}$ & $\langle s \rangle$ & $s_{75}$ & $s_{95}$ & $\langle v \rangle$ & $v_{75}$ & $v_{95}$ & $v_{\rm kin}$ & $\langle Y_e \rangle$ & $Y_{e,75}$ & $Y_{e,95}$ & $\langle d\Omega \rangle_{90}$ \\
	    & $[10^{-1} M_{\odot}]$   & {$[{k_{\rm B}/{\rm baryon}}]$} & $[{k_{\rm B}/{\rm baryon}}]$ & $[{k_{\rm B}/{\rm baryon}}]$ & {[$c$]} & [$c$] & {[$c$]} & {[$c$]} & & & & [$\rm{deg}^{2}$]\\
      \hline
	    \texttt{Q4.chi0.4} & $0.063$ & $14.77$ & $18.23$ & $30.38$ & $0.292$ & $0.311$ & $0.394$ & $0.270$ & $0.048$ & $0.052$ & $0.060$ & $3605.6 $\\
      \texttt{Q4.chi0.6} & $0.285$ & $24.48$ & $34.43$ & $54.68$ & $0.184$ & $0.207$ & $0.276$ & $0.318$ & $0.042$ & $0.044$ & $0.052$ & $2074.7 $\\
      \texttt{Q4.chi0.8} & $0.580$ & $24.85$ & $30.38$ & $62.78$ & $0.201$ & $0.234$ & $0.304$ & $0.348$ & $0.044$ & $0.052$ & $0.056$ & $1957.2 $\\
      \texttt{Q5.chi0.8} & $0.557$ & $17.10$ & $26.33$ & $42.53$ & $0.221$ & $0.241$ & $0.318$ & $0.374$ & $0.043$ & $0.048$ & $0.052$ & $2682.4 $\\
      \texttt{Q6.chi0.8} & $0.376$ & $13.92$ & $18.23$ & $38.48$ & $0.244$ & $0.262$ & $0.331$ & $0.387$ & $0.044$ & $0.048$ & $0.052$ & $3081.9 $\\
      \texttt{Q7.chi0.8} & $0.314$ & $13.47$ & $14.18$ & $22.28$ & $0.258$ & $0.276$ & $0.332$ & $0.399$ & $0.046$ & $0.048$ & $0.052$ & $3333.7 $
    \end{tabular}
  \end{ruledtabular}
  \caption{Bulk thermodynamic and kinematic properties of the dynamical
    ejecta for the various binary configurations. From left to right we
    report: the total dynamical ejecta mass $M_{\rm ej}$ (according to
    the Bernoulli criterion), the mass-weighted entropy per baryon
    $\langle s \rangle$, and the $75$\% and $95$\% percentiles of the
    specific entropy $s_{75}$, $s_{95}$, the average velocity of the
    ejecta $\langle v \rangle$, the corresponding percentiles $v_{75}$,
    $v_{95}$, $v_{\rm kin}$ obtained from the total ejecta energy and
    ejecta mass measurements, with similar diagnostics for the electron
    fraction $\langle Y_e \rangle$, $Y_{e,75}$, $Y_{e,95}$, and, finally
    the $90\%$-confinement region in the ejecta in terms of its angular
    extent $\langle d\Omega\rangle_{90}$. The entropy $s$ is given in
    units of $k_B/{\rm baryon}$, while the velocity measurements are
    given in terms of the speed of light $c$.}
  \label{tab:dyn_ejecta_properties}
\end{table*}

\begin{figure*}
  \centering
  \includegraphics[width=1.0\textwidth]{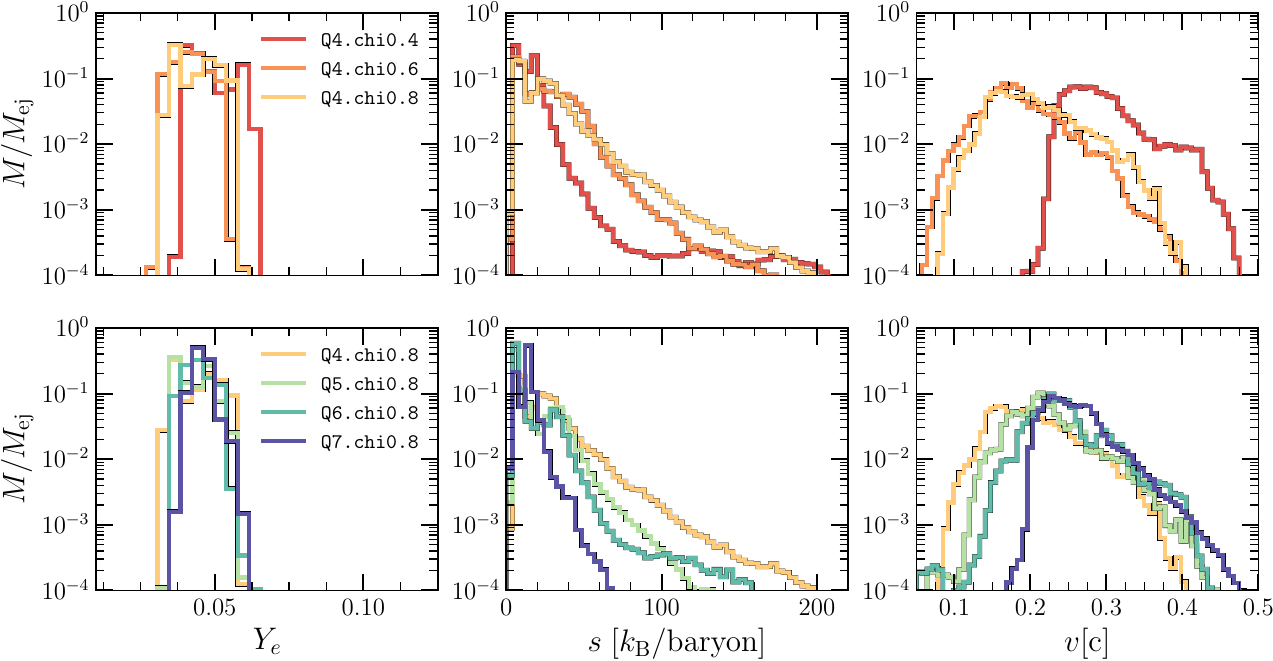}
  \caption{Distributions of the ejected mass in terms of the electron
    fraction $Y_{e}$ (left column), entropy per baryon $s$ (middle
    column), and velocity $v$ (right column), for the six binaries
    considered. In all cases, the data is gathered over the whole history
    of the detector at $r_{\rm det} = 590.8\,{\rm km}$. The collected data was
    binned with chosen ranges: $Y_e\in [0,0.125]$, $s \in
    [0,220]\,[k_{\rm B}/{\rm bar.}]$ and $v / c\in [0.05,0.5]$, for which
    $N=40$, $N=80$ and $N=80$ bins are used, respectively.  }
  \label{fig:ye_s_vel_histogram}
\end{figure*}

Much of this information is summarized and quantified in
Tab.~\ref{tab:dyn_ejecta_properties}, which complements what is presented
graphically in Fig.~\ref{fig:ye_s_vel_histogram}. 

Table~\ref{tab:dyn_ejecta_properties} summarizes the corresponding
mass-weighted means and $75\%$ and $95\%$ percentiles. Thermodynamic and
kinematic distributions are obtained from the time-integrated flux
through the detector at $r_{\rm det}=590.8\,{\rm km}$, considering only
outward-moving ($u^r>0$) matter satisfying the Bernoulli criterion
$hu_t<-1$, and are shown in Fig.~\ref{fig:ye_s_vel_histogram}. The
detector-integrated ejecta mass agrees with the global volume integral to
within $0.5\%$.

Hence, combining the information reported in
Tab.~\ref{tab:dyn_ejecta_properties} and illustrated in
Fig.~\ref{fig:ye_s_vel_histogram}, it becomes clear that strong tidal
disruption causes entropy increase in the ejecta through more prominent
and frequent collisions of the material in the tidal tail, with mean
entropy that is higher than that of weaker tidal disruption
configurations. This is also reflected in the $s_{75}$ and $s_{95}$
percentiles, where a quarter of the material in case of stronger
disruptions achieves values on the average a factor of $2$ higher than
for the remaining cases -- and where $5\%$ of the material features
entropy values greater than $50 k_{\rm B}/\rm{baryon}$.

\subsection{Ejecta bulk velocity}

In terms of velocity, we find that increasing mass asymmetry generically
increases both the values of $\langle v \rangle$ and moves the
percentiles higher in a monotonic fashion with $Q$; this conclusion is
consistent with the findings of Refs.~\cite{Kyutoku2015,
  Hayashi2020}. Although our mass-weighted average velocity of the ejecta
$\langle v \rangle$ is computed differently than the $v_{\rm ave}$
introduced in Ref.~\cite{Kyutoku2015}, the dependence on the mass ratio
and the overall magnitudes agree well. To establish a better
correspondence with the latter diagnostic, we now briefly recall its
definition and compare its values. More specifically, the total energy of
the ejecta is defined as~\cite{Kyutoku2015},
\begin{equation}
  E_{\rm ej} := \int_{u_{t}<-1} (W^2 \rho h - P)\sqrt{\gamma}
  d^{3}x\,.
\end{equation}
The kinetic energy $T_{\rm ej}$ of the ejecta can be subsequently
obtained by subtracting the rest-mass energy from the total energy:
\begin{equation}
  T_{\rm ej} := E_{\rm ej} - M_{\rm ej} - U_{\rm ej}\,,
\end{equation}
and where the third component, the internal energy of the ejecta $U_{\rm
  ej}$, has been neglected due to being an order of magnitude smaller.
The average velocity of the ejecta can be evaluated by assuming Newtonian
dynamics and hence with:
\begin{equation}
  \label{eq:v_kin}
  v_{\rm kin} := \sqrt{\frac{2T_{\rm ej}}{M_{\rm ej}}}\,,
\end{equation}
which is clearly an approximation and does not take into account \eg the
gravitational potential energy. Note that a correction assuming a
Newtonian potential~\cite{Hayashi2020} is not applied to our measurements
and that we have changed the name of this diagnostic quantity to $v_{\rm
  kin}$.

The average velocity of the ejecta computed in this way can be seen in
the last column in the relevant velocity-related tuple in
Tab.~\ref{tab:dyn_ejecta_properties}. In all cases except the least
disruptive merger, $v_{\rm kin} > \langle v \rangle$ (the latter computed
on the basis of the distributions). At the same time, we observe a clear
trend of growing $v_{\rm kin}$ for greater spins at fixed mass
ratio. Similarly, $v_{\rm kin}$ increases when the mass ratio $Q$ is
increased at fixed BH spin. This indicates that the disrupted stellar
material is endowed with significant kinetic energy, amplified by the
presence of additional angular momentum in the system or lesser mass of
the dynamical ejecta where the energy is transferred (for increasing
$Q$).

The latter finding is also captured by the functional description of
$v_{\rm kin}(Q) = (0.01533Q + 0.1907)\,c$ fitted to the data of
~\cite{Kawaguchi2016} in case of irrotational binaries. To capture the
simple observed trend, we have attempted the linear fit $v_{\rm
  kin}(Q,\chi_{_{\rm BH}}) = A + B\,Q + C\,\chi_{_{\rm BH}}$ with
best-fit parameters $A=0.1276$, $B=0.01587$ and $C=0.2034$ on our data;
because of the orthogonality of chosen datasets, the fit is numerically
appropriate but would benefit from additional simulations. At the same time, the
$v_{\rm kin}$ appears to noticeably vary depending on the time of
measurement.

\subsection{Histograms of $Y_e$, $s$ and $v$}

Having provided a general description of the ejecta on the basis of
percentile and mass-weighted mean values, we now describe the
distributions in more detail. One first observation is that the
distribution of the material in terms of the electron fraction is similar
in each case and stays roughly between $Y_{e}\approx 0.04$ and
$Y_{e}\approx 0.06$. This is to be expected for two reasons. 
The narrow distribution around $Y_e\approx0.05$ is partly a consequence
of the absence of neutrino transport in our simulations, such that $Y_e$
is predominantly advected with the fluid. While these values are
consistent with earlier BHNS calculations employing truncated-moment or
leakage treatments~\cite{Kyutoku2018,Brege2018}, more recent
neutrino-transport calculations have shown that neutrino interactions can
protonize a non-negligible fraction of the dynamical ejecta to
$Y_e\gtrsim0.1$~\cite{Hayashi2023,Martineau2024}. Our $Y_e$ distributions
should therefore be interpreted as characteristic of the very
neutron-rich component within the approximations adopted here.

\begin{figure*}
  \centering
  \includegraphics[width=0.48\textwidth]{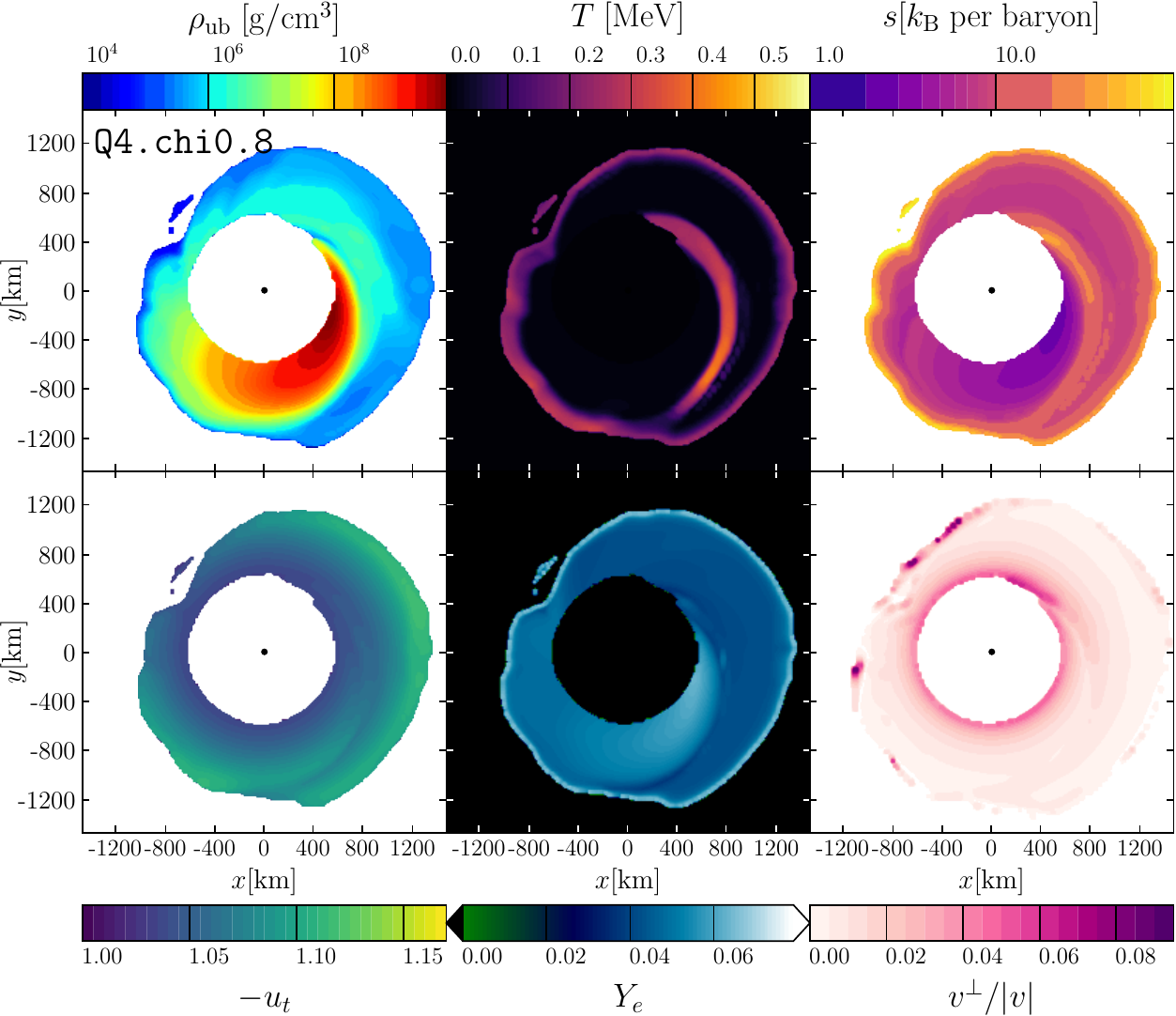}
  \hspace{0.5cm}
  \includegraphics[width=0.48\textwidth]{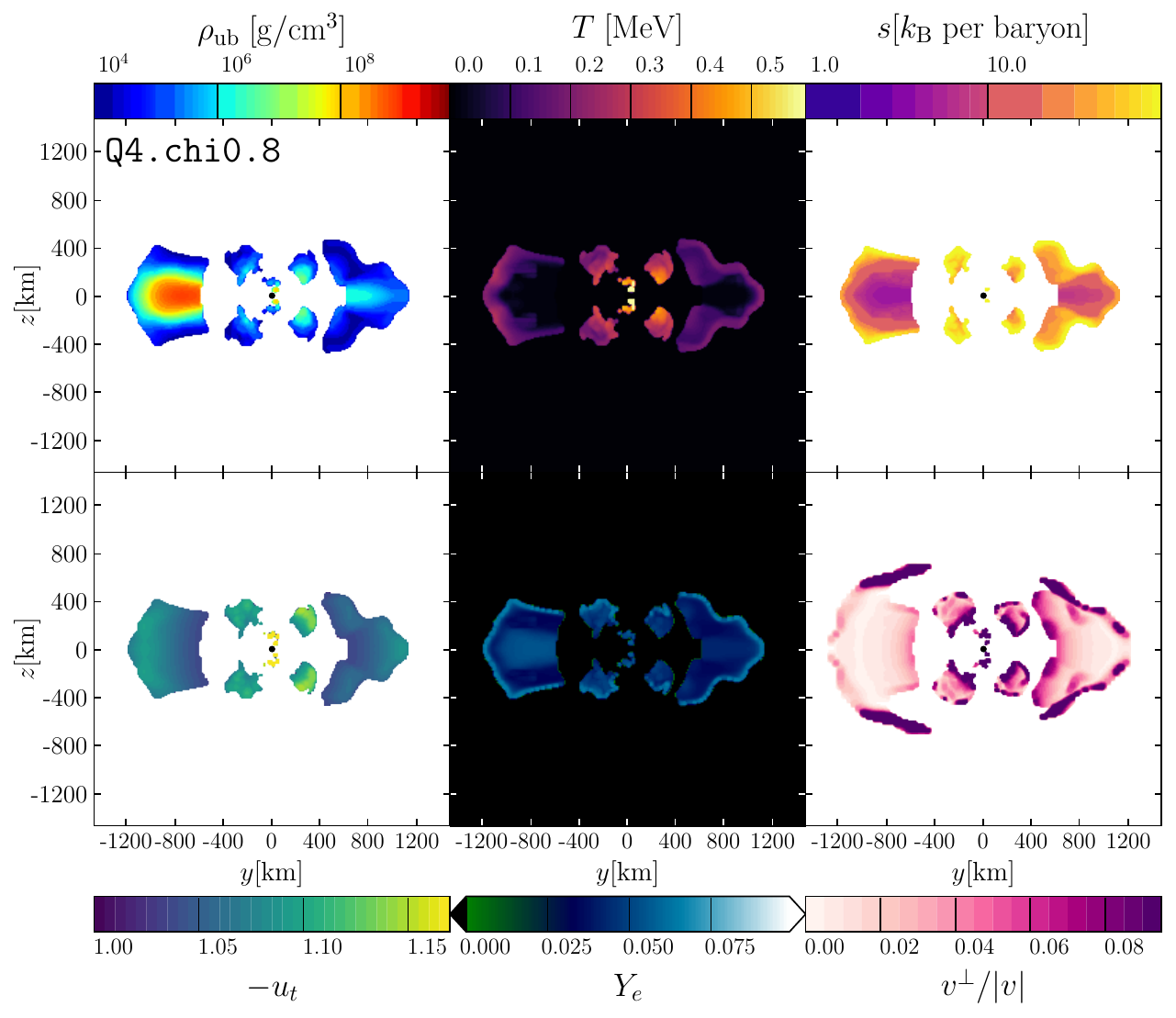}
  \caption{\textit{Left panels:} Snapshots of the dynamical ejecta in the
    equatorial $(x,y)$ plane at the final time of the \texttt{Q4.chi0.8}
    simulation. From the top-left to the bottom-right panel, we report:
    the rest-mass density of the unbound material $\rho_{\rm ub}$, the
    corresponding temperature $T$, the specific entropy $s$, the specific
    energy of the material at infinity $-u_{t}$, the electron fraction
    $Y_e$, and, finally, the ratio between the orthogonal three-velocity
    component and the total three-velocity $v^{\perp}/\lvert v
    \rvert$. \textit{Right panels:} The same as in the right panels but
    when considering sections on the meridional $(y,z)$ plane.}
  \label{fig:Q4.chi0.8_slice_2d_xy}
\end{figure*}

Secondly, we note a large variation in the distribution of the specific
entropy. As anticipated earlier on the basis of mean and percentile
values in Tab.~\ref{tab:dyn_ejecta_properties}, the binaries
\texttt{Q4.chi0.6}, \texttt{Q4.chi0.8} and \texttt{Q5.chi0.8} feature
progressively larger amounts of material with relatively higher values of
entropy. For instance, considering a reference value of $M/M_{\rm
  ej}=10^{-2}$ in Fig.~\ref{fig:ye_s_vel_histogram}, which would refer to
a share of $1\%$ of the material residing in a given bin, the highest
entropy reached would read $s\approx 20\, k_{\rm B}/{\rm baryon}$ for the
\texttt{Q7.chi0.8} and \texttt{Q4.chi0.4} (weak disruption) simulations,
and steadily increase to $s\approx 60\, k_{\rm B}/{\rm baryon}$ for
\texttt{Q4.chi0.8} and \texttt{Q4.chi0.6} cases. The trend of stronger
disruptions leading to larger entropy values in the bulk ($10^{-2}
<M/M_{\rm ej} < 1$) of the material extends all the way to very high
entropy ($s>100\,k_{\rm B}/{\rm baryon}$) values and ``prolongs'' the
tails of respective distributions.

The rightmost panel of Fig.~\ref{fig:ye_s_vel_histogram} reports the
distribution of matter in terms of velocity of the ejecta. Contrasting it
with the preceding panel, even more variance is present in this case,
with the distributions featuring positive skewness, \ie with a milder and
longer fall-off to the right of the average. Note how in the bottom
panel, which reports binaries with $\chi_{_{\rm BH}}=0.8$ and varying
$Q$, a systematic shift of the distribution mean to higher velocity
values is present as the mass ratio increases. When taking into account
the logarithmic scale on the $y$-axis, notable differences are present
when concentrating on the largest mass contributions $10^{-2} < M/M_{\rm
  ej} < 10^{-1}$ in the central region. The systematic shift in ejecta
velocity concerns in particular those bins which contain at least
$5-10\%$ of the material. This behavior anticipates the role of the mean
velocity in determining the characteristics of an EM signal from our
simulations. This is because the mean velocity, along with the opacity
and ejecta mass, sets the photon diffusion timescale $t_{\rm dif} \propto
1/v_{\rm mean}$~\cite{Metzger2017}. The dependence on the velocity is, as
we shall demonstrate shortly, reflected in the variance of our results in
Sec.~\ref{sec:rprocess_nucleosynthesis} and in
Sec.~\ref{sec:radiative_transfer_kilonova}, as it dictates the final
ejecta geometry under homologous expansion. The results for varying spin
at $Q=4$ are visible in the upper right panel, with similar distributions
present for $\chi_{_{\rm BH}}=0.6$ and $\chi_{_{\rm BH}}=0.8$, suggesting
that the transfer of energy and angular momentum (\cf
Sec.~\ref{sec:dyn_mass_EJ}) to the material proceeds in a similar
fashion, with the \texttt{Q4.chi0.8} binary slightly shifted to the
right, as expected from the increased BH spin. The binary
\texttt{Q4.chi0.4} represents an outlier among all our
configurations. While the binary disrupts only very weakly, the sheer
scarcity of the material extends the velocity distribution to higher
values, even featuring a small flat region at $v\approx 0.4\,c$. The
small amount of dynamical ejecta present in this simulation, however,
means that even the largest bins represent no more than $M_{\rm
  bin}\approx 0.07 \times 0.0063\,M_{\odot} \approx 4.4 \times
10^{-4}\,M_{\odot}$ of the material.

\begin{figure*}
  \centering
  \includegraphics[width=0.48\textwidth]{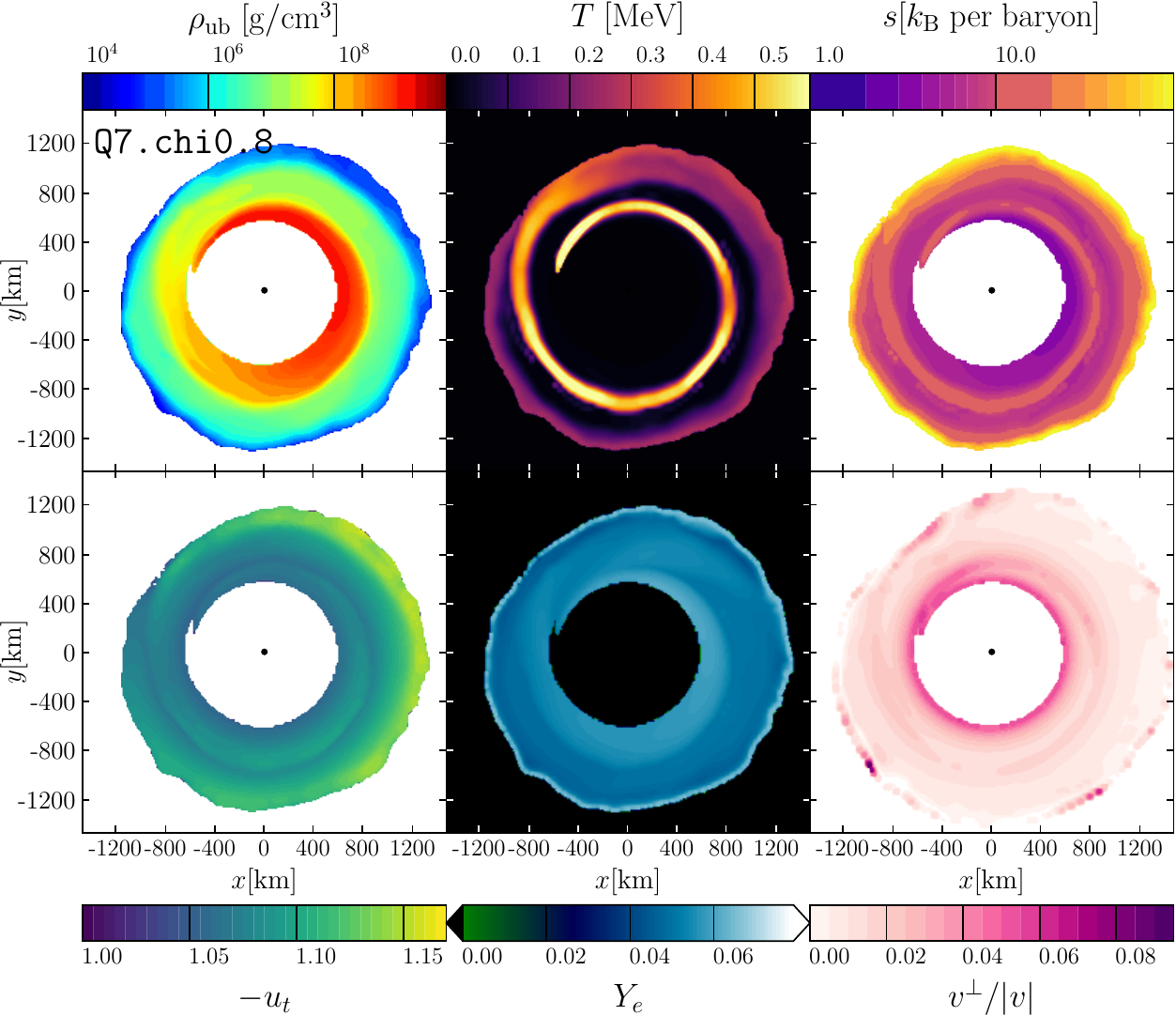}
  \hspace{0.5cm}
  \includegraphics[width=0.48\textwidth]{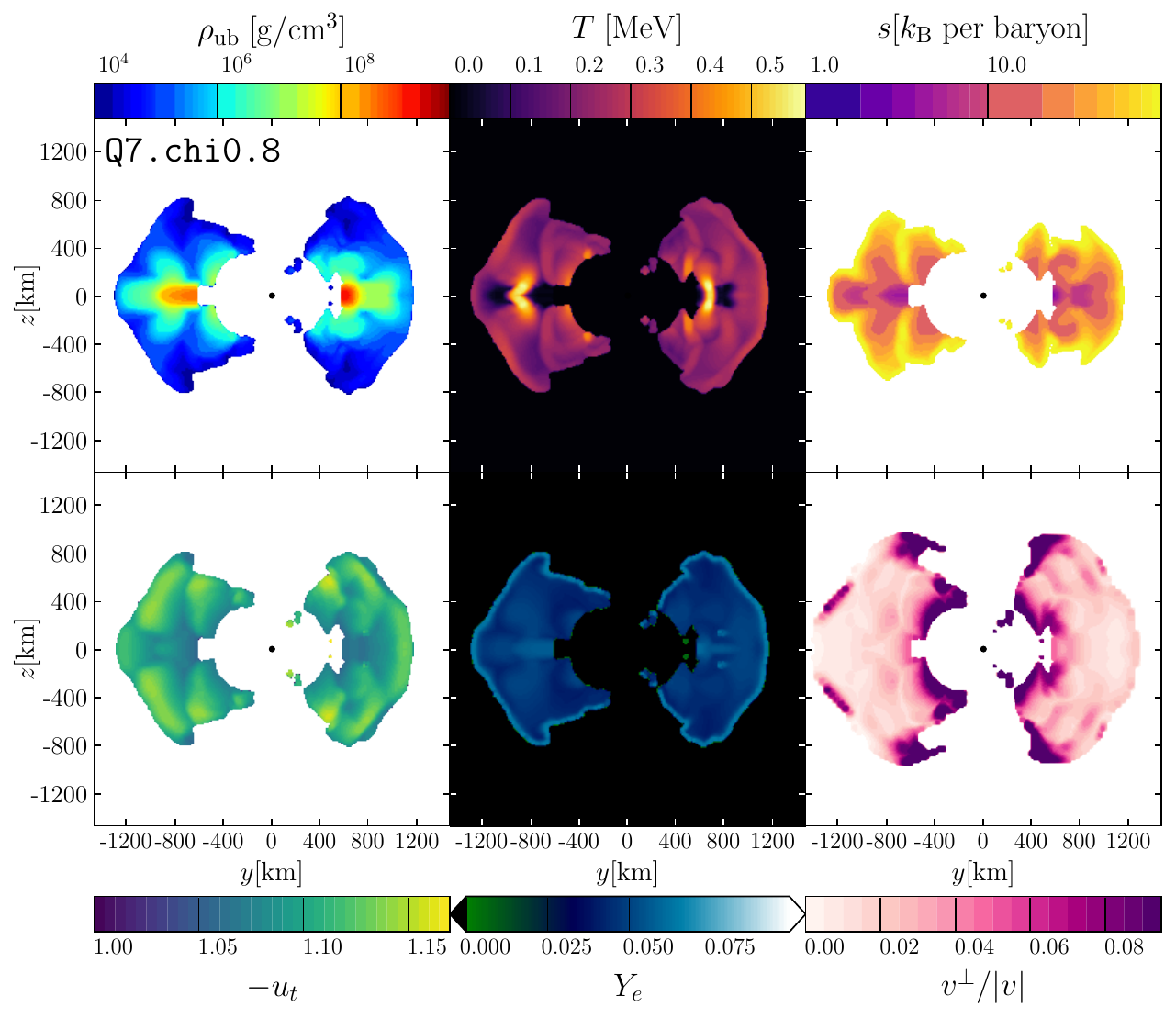}
  \caption{The same as in Fig.~\ref{fig:Q4.chi0.8_slice_2d_xy} but for
    the \texttt{Q7.chi0.8} binary. Note how, in contrast to the
    strong-disruption case of the binary \texttt{Q4.chi0.8}, the ejecta
    for this BHNS system is more axisymmetric.}
  \label{fig:Q7.chi0.8_slice_2d_xy}
\end{figure*}

On a final note, while the high-velocity ($v/c\gtrsim 0.4$) or
high-entropy material ($s \gtrsim 60 k_{\rm B}/{\rm baryon}$) is
responsible for some of the visually most striking differences between
individual plots in Fig.~\ref{fig:ye_s_vel_histogram}, it typically
constitutes a minuscule fraction of total dynamical ejecta
mass. Therefore, its minor contribution to the $r$-process
nucleosynthesis yields can only be appreciated as a variance across
different simulations, but it may be difficult to be captured in terms of
elemental abundances. We note that even the $95\%$-percentiles
(Tab.~\ref{tab:dyn_ejecta_properties}) fall far short of capturing the
extended tails of the respective distributions. However, as we will
discuss in Secs.~\ref{sec:rprocess_nucleosynthesis} and in the
Appendix~\ref{subsec:rprocess_nucleosynthesis_entropy_vel_dep}, the fine
differences in the $r$-process nucleosynthesis yields indicate that the
heavy-elements abundances are crucially correlated with the ``bulk''
features of the distributions, or equivalently, quantified by the
mass-weighted averages $\langle v \rangle$, $\langle s \rangle$ and the
more conservative percentiles $v_{75}$, $s_{75}$. As we will further see
in Sec.~\ref{sec:radiative_transfer_kilonova}, the same conclusions apply
also to the computed kilonova signals discussed therein.

\subsection{Spatial distribution of the ejecta}
\label{sec:geometric_XY_XZ}

The statistical characterization of the ejecta carried out in the
preceding section will now be complemented with a discussion about its
size and variance in space, \ie the extent of the unbound material in the
radial and angular directions, as well as general thermodynamic and
kinematic properties of the ejecta as a function of mass ratio and BH
spin in the various configurations. To that end, in the panels of
Figs.~\ref{fig:Q4.chi0.8_slice_2d_xy}--\ref{fig:Q7.chi0.8_slice_2d_xy} we
present the orbital [\ie $(x,y)$ plane] and meridional [\ie corresponding
  to the $(y,z)$ planes] slices of the system at the end of our
simulations, \ie $t-t_{\rm mer}\approx 15\,\rm{ms}$\footnote{The
simulations at higher mass ratios were carried out for slightly longer to
allow for the settling of the disk. We report the latest common time
across the simulations.}. To save space, we show data only for two
binaries representative of weak and strong tidal disruption, but discuss
the behavior of all simulations in the text.

In each figure, the coordinate axes are set to span $1400\,\mathrm{km}$,
which is the largest range that fully encompasses the dynamical ejecta at
the final time of our simulations. In clockwise order from the upper-left
to the lower-right panels, we show: the rest-mass density of the unbound
material $\rho_{\rm ub}$, the temperature $T$, the entropy per baryon
$s$, the ratio of the angular velocity component (constructed
from $v^\theta$ and $v^\varphi$) to the total velocity magnitude,
$v^{\perp}/|v|$, the electron fraction $Y_e$, and finally the specific energy at infinity $-u_t$. 

We first discuss the properties of the ejecta in the orbital plane, which
are shown in the set of six panels on the left of
Figs.~\ref{fig:Q4.chi0.8_slice_2d_xy}--\ref{fig:Q7.chi0.8_slice_2d_xy}. At
the time when the two-dimensional (2D) slices are extracted, the general
density pattern of the unbound material has a crescent-like shape
reflective of its history, \ie with a prominent high-density region of
varying azimuthal extent (red and orange colours) -- for the
\texttt{Q4.chi0.6} configuration (not shown here), the approximate
azimuthal extent of this region is $150^{\circ}$, and similarly for
\texttt{Q4.chi0.8}. In \texttt{Q5.chi0.8} (also not shown), the angular
extent of the orange-red region covers over $180^{\circ}$ around the
remnant already, while for the \texttt{Q6.chi0.8} and \texttt{Q7.chi0.8}
simulations, it amounts to roughly $240^{\circ}$ and nearly
$360^{\circ}$, respectively. The higher-density part of the
\texttt{Q4.chi0.4} simulation also extends over almost the entire
azimuthal range. Clearly, the azimuthal extent here is directly related
to the duration of the tidal disruption phase; the tidal tail wraps
around the BH more times for higher mass ratios, thus redistributing the
high-density material more evenly. The orange-red regions are also
thinner for higher mass ratios $Q=6,7$ than in the case of smaller mass
asymmetries. Note that this distribution is subject to change over time
as the ejecta expands (the velocity is not yet purely radial), but should
give a reasonable estimate. Clearly, maximal rest-mass densities in the
ejecta occur for the strongest tidal disruption cases --
\texttt{Q4.chi0.8} and \texttt{Q5.chi0.8}, in which case the rest-mass
density reaches $\rho_{\rm ub}\approx 10^{9}\,\rm{g}/\rm{cm}^{3}$ in the
densest part of the tidal tail. These maximum densities drop by a factor
of several when the initial BH spin is smaller or, alternatively, when
the mass ratio of the system is increased. The entirety of the dynamical
ejecta at inspection time is at a distance of at least $600\,\rm{km}$
from the remnant BH, but typically no further than $1300\,\rm{km}$,
depending on the case, and with a variance in the radial extent dictated
by the preferential direction due to the tidal disruption. For instance,
the binary \texttt{Q4.chi0.8} features material which is less than
$100\,\rm{km}$ wide at the narrowest point and spans $700\,\rm{km}$ in
the radial direction at maximum width, whereas the \texttt{Q7.chi0.8}
binary is much more uniform in the width of ``ejecta ring'', with the
shell width between $450$ and $650\,\rm{km}$. We note that the asymmetric
distribution of material in the case of \texttt{Q4.chi0.8} closely
resembles a strong disruption case presented in Fig.1 of
~\cite{Bulla2020}.

Regarding the temperature distribution, the intermediate-density unbound
material ($10^{6}$-$10^{7}\mathrm{g/cm^3}$) located between the outermost
and innermost layers of the dynamical ejecta is predominantly cold ($T
\lesssim 0.1\mathrm{MeV}$). We note that the version of the DD2 EOS we
are using~\cite{Stellarcollapse} extends all the way down to
$0.01\,\rm{MeV}$. The same comment applies to the high-density regions
mentioned above. In contrast, the outer ejecta layers with $T \approx
0.1$-$0.2\,\mathrm{MeV}$ correspond to the tidal tail segment that wraps
around the remnant, leading to a temperature increase upon interaction
with other parts of the tail. The outermost portion of this material,
however, interacts with the atmosphere, which might bias its measured
temperature. A higher-temperature pattern ($T \gtrsim 0.4\,\mathrm{MeV}$)
extends along the tidal tail in a spiral-like fashion, connecting the
outer ejecta layers with the inner high-density region. This hotter
region, prominently present in Fig.~\ref{fig:Q4.chi0.8_slice_2d_xy} for
\texttt{Q4.chi0.8} lies precisely at the interface between the
high-density core and the surrounding intermediate-density layers, thus
neighbouring both. Given that the \texttt{Q7.chi0.8} model exhibits
stronger self-wrapping of the tidal tail at the inspected time, the
resulting high-temperature spiral structure is correspondingly even more
apparent and appears thinner. Clearly, the entropy in the ejecta follows
closely the distribution of temperature and similar comments regarding
its spatial distribution apply, with higher entropy regions concentrating
in the parts of the tidal tail subject to prior collisions, or
interactions with the atmosphere. The former represent a considerable
part of the ejecta -- especially for \texttt{Q4.chi0.8} and
\texttt{Q5.chi0.8} configurations -- and are reflected in the densely
populated high-entropy bins in Fig.~\ref{fig:ye_s_vel_histogram}. On the
other hand, the tails of the entropy distribution are the result of
interactions of the ejected material with the atmosphere.

The specific energy at infinity $-u_{t}$ can be interpreted as the
asymptotic Lorentz factor measured by an observer at
infinity\footnote{This is because $\lim\limits_{r\to \infty} W =
\lim\limits_{r\to \infty} \alpha u^{t} = \lim\limits_{r\to \infty}
g^{t\nu}u_{\nu} = -u_{t}$.} and is hence indicative of the local velocity
of the ejecta. We note that the distribution of $-u_{t}$ is more uniform
across the ejecta in the equatorial plane, with a profile that is
increasing in the direction of outer ejecta layers. Furthermore,
significant anisotropy is present in the distribution of highly unbound
material with $-u_{t} \gtrsim 1.1$ -- easily observed, for example, in
one-half ($ -180^{\circ}< \varphi < 180^{\circ}$) of the azimuthal range
in Fig.~\ref{fig:Q7.chi0.8_slice_2d_xy}. As already commented before and
shown in
Figs.~\ref{fig:Q4.chi0.8_slice_2d_xy}--\ref{fig:Q7.chi0.8_slice_2d_xy},
the electron fraction is uniformly distributed and typically around
$Y_{e}\approx 0.05$ in the equatorial plane - its slightly higher values
indicated by a lighter shade of blue correlate with the over-density
regions represented by orange-red parts of the colormap (\cf first panel
of each combined figure).

Finally, also shown is the ratio between the magnitude of the non-radial
(orthogonal) three-velocity component, \hbox{$\bm{v}^{\perp} := \bm{v} -
  v^{r}\partial_r$}, and the total three-velocity magnitude,
\hbox{$\lvert v\rvert := (\gamma_{ij}v^{i}v^{j})^{1/2}$}. This ratio
quantifies the relative contribution of the non-radial motion to the
overall three-velocity of the material, and at the same time serves an
auxiliary role in assessing the quality of the data extraction and
subsequent post-processing. Namely, for the hand-off of the simulation
data to a radiative-transfer code, the ejecta material is expected to
reach homologous expansion~\cite{Roberts2011, Rosswog2014a, Grossman2014,
  Neuweiler2022, Collins2022, Shingles2023, SippensGroenewegen2025,
  Magistrelli2026}; additionally, the non-radial velocity components
ought to be sufficiently small so that no significant shell crossing or
compositional mixing occurs over the radiative-transfer
timescale. Although the non-radial velocity component constitutes a
non-negligible contribution to the total velocity in all of our
simulations, its magnitude even in the densest ejecta region is smaller
than about $5\%$ in the equatorial plane -- where the material is most
dense -- and otherwise smaller than $10\%$ in the regions away from the
orbital plane (see also the meridional view in the right panels of
Fig.~\ref{fig:Q7.chi0.8_slice_2d_xy}). In the remaining, more tenuous
part of the ejecta, the non-radial contribution drops to approximately
$2\%$ and remains relatively small regardless of the spatial location.

Similar conclusions regarding the spatial distribution of the ejecta can
be made when considering a meridional $(y,z)$ plane and bearing in mind
that we impose the reflection symmetry in the ${\rm z}$-direction to
reduce the computational costs and simplify our analysis of the
direction-dependent bolometric luminosities in
Sec.~\ref{sec:radiative_transfer_kilonova}. In this case, and
concentrating on the set of six panels on the right of
Figs.~\ref{fig:Q4.chi0.8_slice_2d_xy}--\ref{fig:Q7.chi0.8_slice_2d_xy},
the large degree of (azimuthal) asymmetry in the distribution of the
rest-mass density is even more apparent for strongly disruptive mergers
such as~\texttt{Q4.chi0.8},~\texttt{Q5.chi0.8} and \texttt{Q4.chi0.6}
(see, \eg right panels of Fig.~\ref{fig:Q4.chi0.8_slice_2d_xy}), the
degree of which subsides for $Q\geq 6$, and where the directional
variance is not as strongly pronounced (see right panels of
Fig.~\ref{fig:Q7.chi0.8_slice_2d_xy}).

Notwithstanding the complex and varied phenomenology of the spatial
distribution of the ejecta described so far, some simple and robust
features can be easily recognised and are summarized briefly below:
\begin{itemize}
\item tidal disruption at smaller mass ratios features large directional
  asymmetry in the ejecta profile, as well as on average higher
  temperatures and entropy,
\item binaries with larger mass asymmetry display a greater abundance of
  high-velocity ejecta, and a more radially extended distribution,
%
\item stronger tidal disruptions at lower mass ratios produce a more
      directionally asymmetric and azimuthally confined ejecta
	distribution, while weaker disruptions exhibit a broader
		and more nearly axisymmetric azimuthal extent.
\end{itemize} 

\subsection{Angular distribution of the ejecta}
\label{sec:ejecta_angular_distribution}

Having discussed the dynamics of matter in the strong-field
region, we now turn to studying the angular distribution
of the ejecta as this has a direct impact on the measurement made by a
distant observer. This can be done by inspecting the flux of matter
through a suitably chosen detector surface and hence we measure the
surface flux of the unbound matter ($\Phi_{\rm ub}$) with the following
integral:
\begin{equation}
	\frac{dM_{\rm b, ub}}{dt} = \Phi_{\rm ub} :=
        \oint_{S} \sqrt{-g} W \rho u^{r}~ d^{2}x^{A}\,,
\end{equation} 
where only fluid elements with $u^{r}>0$ (outward directed flow) and $h
u_{t}<-1$ are considered, and with \hbox{$d^{2}x^{A} := r^{2}\sin\theta
  d\theta d\varphi$} the standard spherical infinitesimal surface
area. The computation of surface integrals and the production of surface
data output in the \texttt{HEALPix} format~\cite{Gorski2005} have been
very useful in a recent investigation of the accretion of fall-back
material from a binary neutron-star merger~\cite{Musolino2024}. We here
use a resolution of $N_{\rm side}=32$, which gives us a total of $N_{\rm
  pix} = 12 N_{\rm side}^{2} = 12288$ cells in the \texttt{HEALpix} grid.

\begin{figure}
  \centering
  \includegraphics[width=0.99\columnwidth,height=0.44\textwidth]{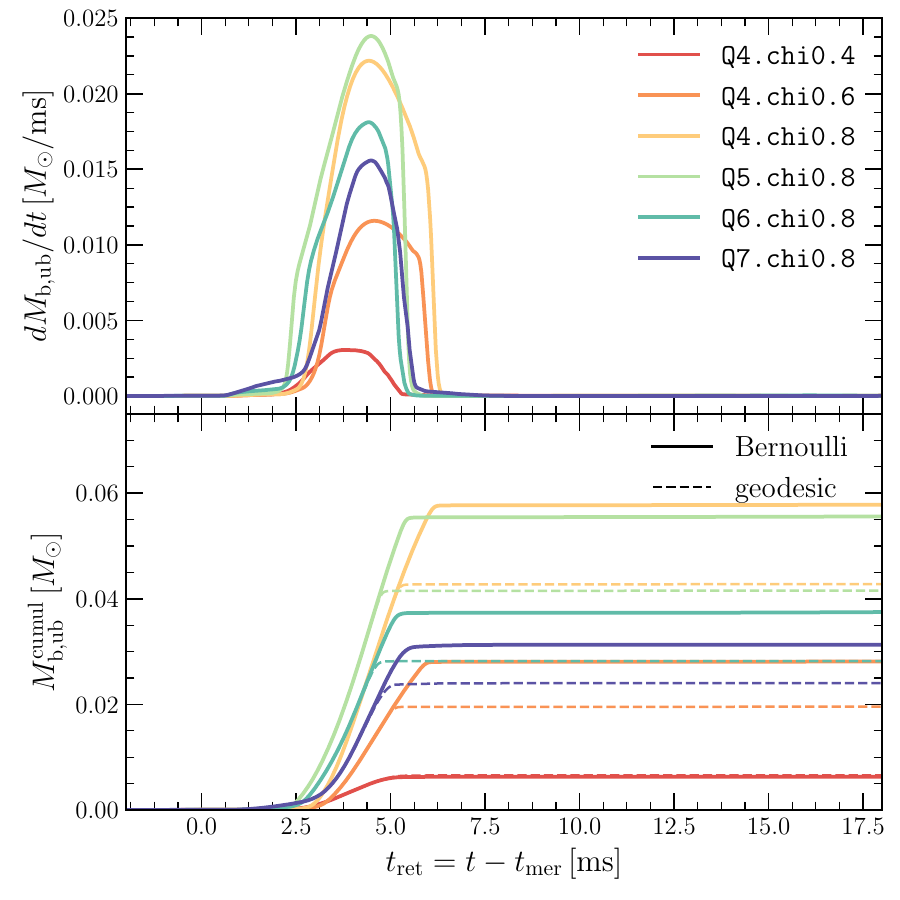}
  \caption{ \textit{(Top panel)} Flux of the unbound mass component
    through a detector at $r = 300\, M_{\odot}$ as a function of retarded
    time, measured according to the Bernoulli criterion. Note how in all
    simulations the large majority of the ejected mass is recorded over a
    window of $\approx 5\,\rm{ms}$. The amplitude of the peak is directly
    correlated with the strength of the disruption, so that a WTD leads
    to a small mass flux and, viceversa, a STD leads to a large mass
    flux. \textit{(Bottom panel)} Integrated unbound mass flux through
    the detector as a cumulative measure, according to the Bernoulli
    (solid lines) and geodesic (dashed lines) unboundness criteria. For
    particularly strong disruptions (\eg \texttt{Q4.chi0.8}), Bernoulli
    criterion yields $50\%$ larger measurements of the unbound mass than
    the geodesic criterion.}
  \label{fig:mass_flux}
\end{figure}

The corresponding mass-ejection fluxes and their time integrals (denoted
as $M_{\rm b,ub}^{\rm cumul}$) are shown in Fig.~\ref{fig:mass_flux} for
a detector at a fiducial distance of $300\,M_{\odot}$. For completeness,
we present the cumulative measures according to both the Bernoulli and
the geodesic unboundness criteria. Interestingly, independent of the
mass ratio and BH spin, the dynamical ejecta reaches the referential
detector position at roughly the same time after the merger $t_{\rm
  ret}:= t-t_{\rm mer} \approx 2\,\rm{ms}$, with the flux momentarily
peaking about $2\,\rm{ms}$ after the first unbound fluid elements have
crossed the surface. The crossing of the detector surface ends at about
$t_{\rm ret}\approx6\,\rm{ms}$, implying that the entirety of the
dynamical ejecta has moved past a distance of $300\,M_{\odot}$ by this
time. Additionally, we have checked that at a time $t_{\rm ret}\approx
25\,\rm{ms}$, over $90\%$ of the total dynamical ejecta mass in each case
has crossed the farthest detector surface placed at
$r=750\,M_{\odot}\approx1107.75\,\rm{km}$.

Given that the numerical grid extends to $1024\,M_{\odot}$ and the
resolution (and hence the accuracy with which the properties and mass of
the dynamical ejecta are measured\footnote{While the rest-mass outside of
the BH is conserved thanks to the flux-conservative formulation of the
general-relativistic magnetohydrodynamics (GRMHD) equations, the
identification of unbound elements will degrade as the resolution
decreases radially outwards.}) drops at large radii, a compromise must be
made regarding the location of the detector where the data is
collected. In the previous sections, we have decided to measure and
report the bulk properties of the ejecta based on the data collected at
$400\,M_{\odot}\approx 591\,\rm{km}$, in particular, when considering the
distribution of the various fluid properties. This is also the radius at
which tracers for $r$-process nucleosynthesis in
Sec.~\ref{sec:rprocess_nucleosynthesis} are extracted. On the other hand,
when considering the data useful for the radiative-transfer simulations
reported in Sec.~\ref{sec:radiative_transfer_kilonova}, we will use
three-dimensional snapshots extracted at the latest possible time, whose
2D projections will be discussed in
Sec.~\ref{sec:ejecta_angular_distribution}.

Regardless of the location of the detector, the evolution of the mass
flux shown in Fig.~\ref{fig:mass_flux} is quite similar in all cases,
with larger magnitudes clearly corresponding to stronger tidal disruption
cases. The same could also be demonstrated for other detector
radii. While the presentation of this flux does not necessarily bring new
information to the quantitative analysis, it informs the choice of the
detector for extracting artificial tracers in
Sec.~\ref{sec:rprocess_nucleosynthesis}.

\begin{figure*}
  \centering
  \includegraphics[width=1.0\textwidth]{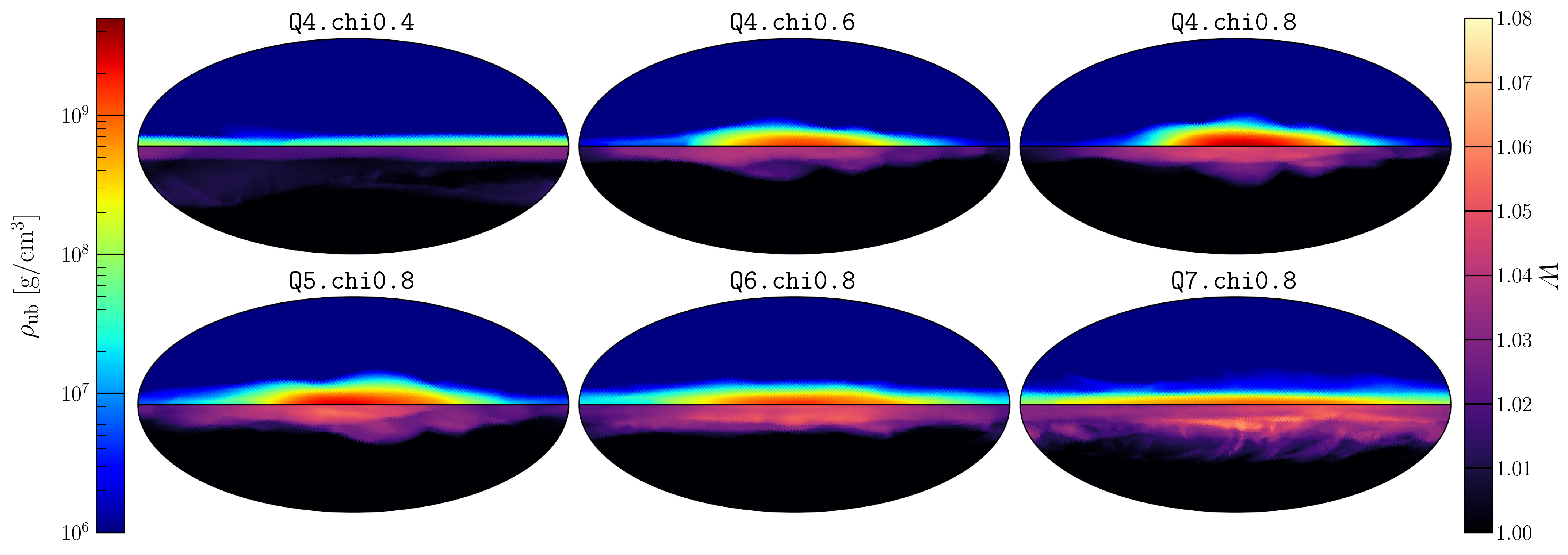}
  \caption{Angular distribution of the rest-mass density of the unbound
    matter (upper hemispheres) and of the corresponding Lorentz factor
    (lower hemispheres), averaged over the active history of the detector
    at $r = 400\,M_{\odot}$. Weak tidal disruptions (\texttt{Q4.chi0.4},
    \texttt{Q6.chi0.8}, \texttt{Q7.chi0.8}) evidently display a higher
    degree of axisymmetry and uniformity in the distribution, contrasting
    with the large directional dependence of strong tidal disruptions.
    The maps have been rotated so that the mass-weighted mean of the
    distribution is located at a longitude of $\varphi=0^{\circ}$.}
  \label{fig:cumulative_ang_dist_healpix_all}
\end{figure*}

To complement the information about the surface flux given
Fig.~\ref{fig:mass_flux}, we report in
Fig.~\ref{fig:cumulative_ang_dist_healpix_all} the geometrical
distribution of some of the properties of the ejected mass for the six
binaries considered. More specifically, we present the time-integrated
distributions across the entire surface of the detector of the rest-mass
density of the unbound material $\rho_{\rm ub}$ (northern parts of each
panel) and of the corresponding Lorentz factor $W$ (southern
parts). Because the distributions are highly anisotropic and the degree
of anisotropy differs among the various binaries, the sky maps have been
shifted so that the center of the \texttt{HEALpix} grid (at
$\varphi=0^{\circ}$) corresponds in each case to the location of the
mass-weighted mean of the angular distribution; this facilitates the
comparison across different scenarios. It is then apparent from
Fig.~\ref{fig:cumulative_ang_dist_healpix_all} that ``weak'' binary
mergers (\cf Tab.~\ref{tab:ID_properties}) feature, on average, a higher
degree of axisymmetry in the ejecta, evidenced by the polar and azimuthal
extent in the top panels, in addition to higher velocities, represented
by larger (here, brighter) Lorentz factors. On the other hand, ``strong''
tidal disruptions feature both a higher degree of collimation in the
dynamical ejecta and, at the same time, slightly lower Lorentz
factors. In this sense, the cumulative spatial picture corresponds
closely to the findings made in reference to the distributions presented
in Fig.~\ref{fig:ye_s_vel_histogram}.

\begin{figure}
  \centering
  \includegraphics[width=1.0\columnwidth]{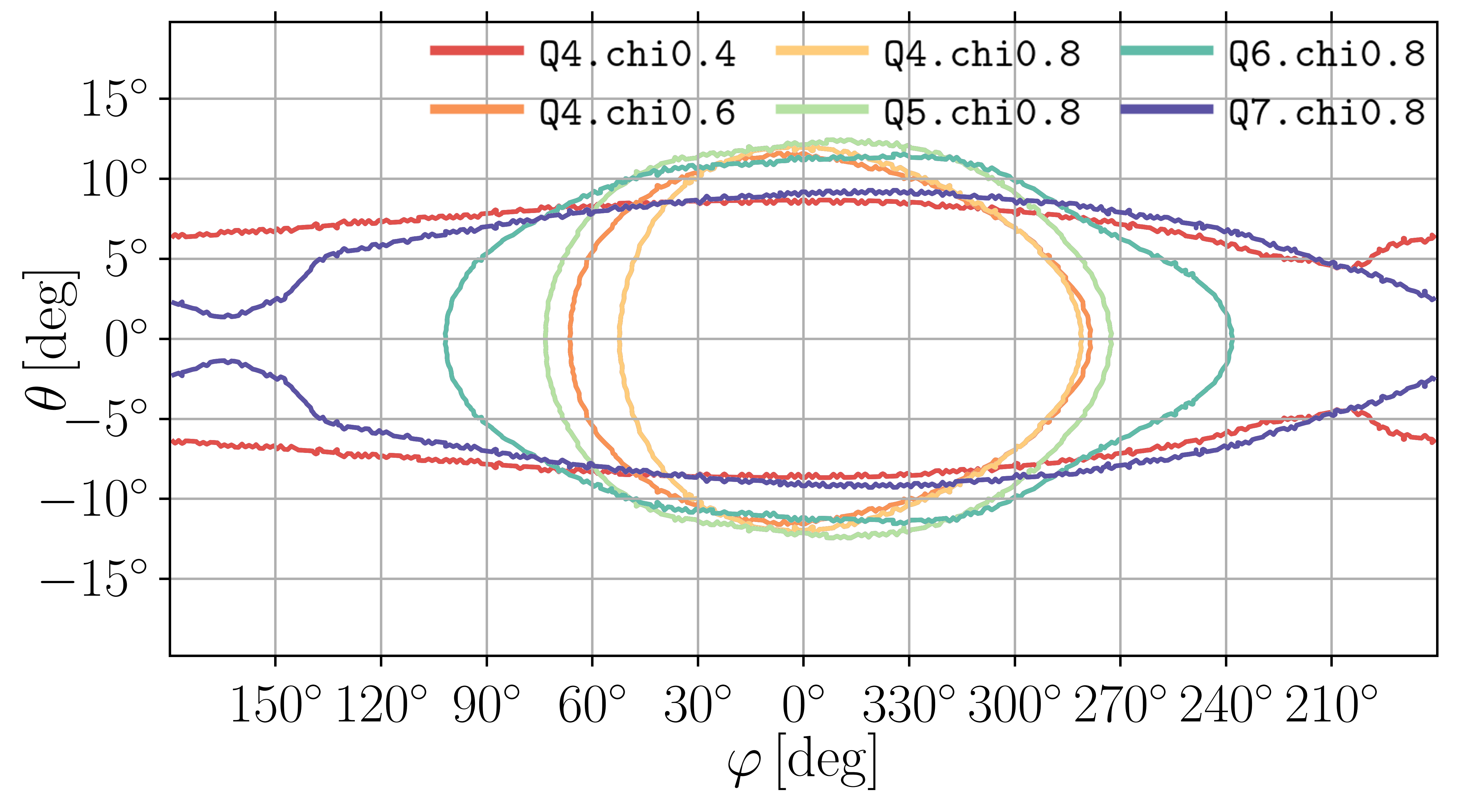}
  \caption{Low-latitude view of the angular extent of the dynamical
    ejecta for the various binaries. Shown are the $90\%$-confinement
    regions, \ie the area of the sky, where $90\%$ of the dynamical
    ejecta mass in a given simulation has been recorded by a given
    detector. Note that such an area spans between $8\%$ of the sky for
    weakly disruptive mergers \texttt{Q4.chi0.4} and \texttt{Q7.chi0.8}
    and $4-5\%$ for the strong tidal disruptions \texttt{Q4.chi0.8} and
    \texttt{Q5.chi0.8}. 
    }
  \label{fig:hpmoc_maps_angular_extent}
\end{figure}

The properties of the angular extent can be made more quantitative by
reporting in Fig.~\ref{fig:hpmoc_maps_angular_extent} the $90\%$
''confinement regions'' of the dynamical ejecta, that is, the contours
within which $90\%$ of the total ejected mass is contained. To obtain
such a representation -- which, to the best of our knowledge, has not
been reported before -- we first convert the 2D array of values of the
unbound rest-mass density into a one-dimensional (1D) array. Then, the
array is sorted according to the largest values, which are given the
highest priority and thus correspond always to the densest regions in the
ejecta. Finally, a sum is performed over the locations of all the pixels
in the spherical grid, until $90\%$ of the total ejecta mass is reached,
\ie when $\sum_{i_{\rm pix}} \rho_{\rm ub}(i_{\rm pix}) W(i_{\rm pix})
dA_{i_{\rm pix}} = 0.9 M_{\rm ej}$. Here, the discrete surface area
element, consistently with the equal-size \texttt{HEALpix} convention, is
equal to $4\pi / N_{\rm pix}$. A rapid inspection of the figure reveals
that the method described above is robust, as the fall-off of the ejecta
rest-mass density projected on the detector surface is smooth and not
subject to large gradients (this would lead to clear zig-zag contours).

A few comments regarding the found angular extent are in order. First,
for all of the simulations, the total polar extent of the ejecta is
$\lesssim 20^{\circ}$ across both hemispheres, which is characteristic
for the purely dynamical ejecta component (this is similar to \eg
Fig.~$5$ of~\cite{Foucart2017b}, where most of the material is contained
within $-0.1 \leq \cos\theta \leq 0.1$). Second, weak tidal disruptions
are narrower in the polar direction -- averaging around
$5^{\circ}-7^{\circ}$ on each side; at the same time, the
$90\%$-confinement regions are much more uniform in the azimuthal
direction. Finally, the opposite trend can be observed for strongly
disruptive mergers, since for such binaries the ejecta tends to be more
collimated, and its azimuthal extent averages around $120^{\circ}$
instead of the entire azimuthal range. We have computed the total angular
area occupied by ejecta within these contours (\cf the last column of
Tab.~\ref{tab:dyn_ejecta_properties}), and find that it ranges between
$4\%$ of the entire sky for strong tidal disruptions and $8\%$ for weak
tidal disruption cases. The polar opening angle of the largely equatorial
dynamical ejecta clearly depends on the strength of disruption and ought
to be reflected in kilonova models for BHNS mergers. Its impact on the
signal has been studied systematically in a limited fashion (see, \eg
~\cite{Darbha2020}), with most models fixing a single ejecta geometry and
varying other parameters. State-of-the-art approaches for BHNS systems
often rely on fixed ejecta geometries, either following a well-motivated,
parametrized ansatz,~\eg~\cite{Kawaguchi2020c, Anand2021} or import
simulation data directly~\cite{Tanaka2013b, Bulla2020, Markin2026}. We
believe that the detailed description of the angular and in particular
polar extent provided in this paper can be useful for future models of
ever more refined synthetic BHNS kilonova light-curves.

\section{$r$-process nucleosynthesis}
\label{sec:rprocess_nucleosynthesis}

We next present the nucleosynthetic yields obtained by employing the
\texttt{SkyNet}~\cite{Lippuner2015} nuclear-reaction network on
artificial tracers probing the composition and thermodynamical state of
dynamical ejecta in our simulations. We recall that \texttt{SkyNet} is a
general-purpose nuclear reaction network written in C++, with a highly
modular structure. In addition to governing the evolution of elemental
abundances $Y_i := n_i / n_{\rm B}$ (with $n_{\rm B}$ the baryon number
density and $n_i$ the number density of a given nuclear species) through
strong~\cite{Cyburt2010, Frankel1947, Mamdouh2000, Wahl2002, Panov2010},
weak~\cite{Fuller82, Oda1994, Langanke00, Cyburt2010}, and inverse
nuclear reactions, the code contains a nuclear statistical equilibrium
solver, as well as support for the Helmholtz EOS governing the
thermodynamics~\cite{Timmes2000}. In addition, the nuclear-reaction
network in \texttt{SkyNet} is a self-heating network, that is, the
nuclear reactions and decays are reflected in the temperature evolution
along thermodynamic trajectories by book-keeping entropy changes. Other
publicly accessible nuclear-reaction networks are available, \eg
\texttt{WinNet}~\cite{Reichert2023}, and have found wide application in a
variety of astrophysical contexts~\cite{Winteler2012, Korobkin2012,
  Bovard2017}.

\subsection{Thermodynamic trajectories and tracer extraction}

Following Refs.~\cite{Radice2016,Bovard2016,Radice2018a,Ng2024c}, we
construct artificial tracers from the unbound material crossing the
detector. The $(s,Y_e)$ data are binned on a $120\times120$ grid, and
tracers are sampled from the nonempty bins according to their accumulated
mass, yielding $\sim4{-}4.5\times10^4$ tracers per model. These weights
are subsequently used to combine the \texttt{SkyNet} yields.

We assume that the nuclear statistical equilibrium (NSE) is the starting
point for each tracer trajectory. To that end, the artificial tracer
collection needs to be slightly modified in order to reflect the
evolution of a tracer from such a point in time. We further assume that
the thermodynamical state of the ejecta at those times is described by
the modified Timmes EOS~\cite{Timmes2000} and retains the original
entropy $s$ and electron fraction of the tracer $Y_{e}$. Subsequently, a
referential temperature of $T=6\,\rm{GK}\approx 0.517\,\rm{MeV}$
corresponding to NSE conditions is imposed~\cite{Lippuner2015} (the last one is typically
higher than the actual temperature in our ejecta) and a new, initial
rest-mass density $\rho_{0}$ of the expanding material is recovered from
the EOS. Finally, the dynamical timescale of the expansion of ejecta
$\tau$ is obtained by setting the $r$-process related density evolution
\begin{equation}
	\rho(t) = \rho_{0}\left(\frac{3\tau}{e\, t}\right)^{3}\,,
	\label{eq:rproc_dens_evol}
\end{equation} 
equal to the homologous evolution of density 
\begin{equation} 
  \rho(t) = \rho_{\rm ext}\left(v_{\rm ext} t / r_{\rm
    ext}\right)^{-3}\,,
\end{equation}
where $\rho_{\rm ext}$ and $v_{\rm ext}$ are the density and velocity of
the fluid element when it crosses the detector surface at $r_{\rm ext}$
at time $t$, respectively.

\begin{figure}
  \centering
  \includegraphics[width=1.0\columnwidth]{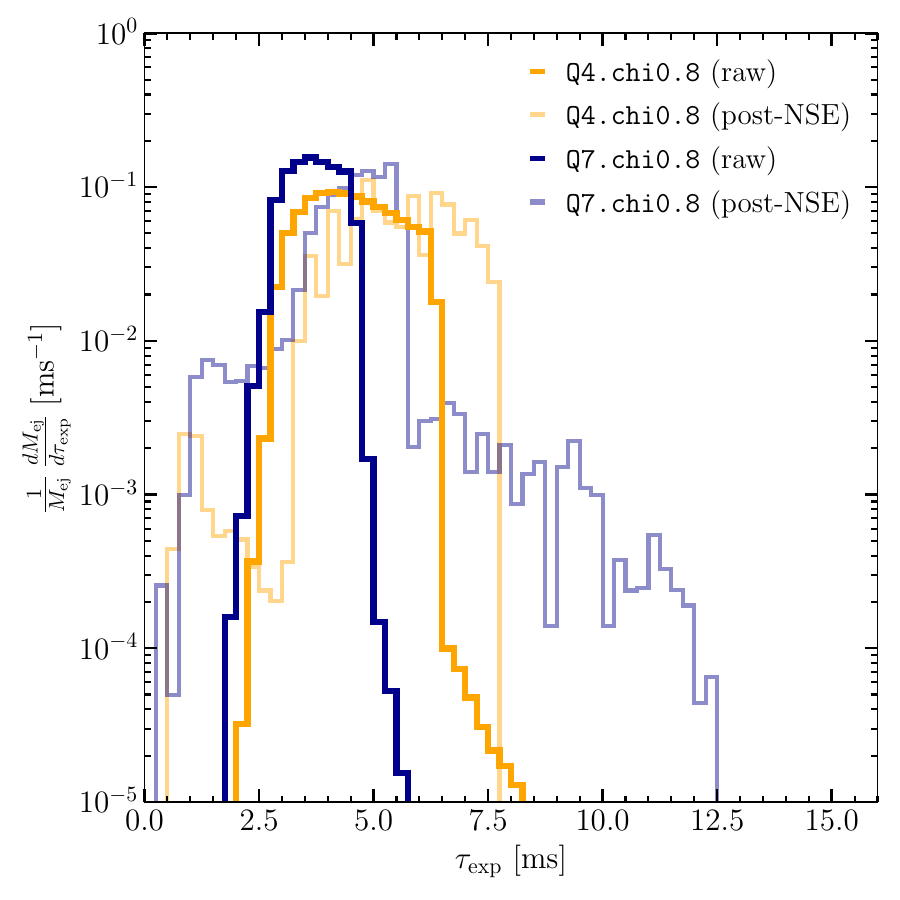}
  \caption{The distributions of ``bare'' and NSE-adjusted expansion
    timescales $\tau$ for tracer trajectories recovered from the history
    of unbound material passing a detector at $r=400M_{\odot}$. A
    strong-disruption \texttt{Q4.chi0.8} and a weak-disruption
    candidate~\texttt{Q7.chi0.8} are contrasted, illustrating the inverse
    relationship between the ejecta velocity and computed expansion
    timescales. The expansion timescales that enter the nuclear reaction
    network~\texttt{SkyNet} are spread between $10^{-2}~\rm{ms}$ and
    $15~\rm{ms}$.}
  \label{fig:expansion_timescales}
\end{figure}

Accordingly, the ``bare'' expansion timescales $\tau_{\rm bare}$ inferred
from the velocity of the material measured on the detector surface are in
general not equal to the ones entering the nuclear reaction network, but
are instead related by $\tau = \tau_{\rm bare} ~ \left(\rho
/\rho_{0}(Y_{e}, s) \right)^{3}$\footnote{This relation follows by
setting the right-hand sides of Eq.~\eqref{eq:rproc_dens_evol} equal,
with two pairs of $\tau$ and initial $\rho$.}, with density at NSE
depending on the
entropy and composition. We present the distributions of both of these
quantities for inspected tracer families in
Fig.~\ref{fig:expansion_timescales}, for a strong-disruption binary
\texttt{Q4.chi0.8} and a weak-disruption~\texttt{Q7.chi0.8}. The
uncorrected expansion timescales are plotted with solid lines, while
those after the correction are semi-transparent. Note that correcting for
NSE shifts both distributions toward slightly longer timescales and
increases by orders of magnitude the amount of tracers for which
$\tau\lesssim~2\rm{ms}$. However, in either case the mapping between
$\tau$ and observed ejecta velocity in histograms of
Fig.~\ref{fig:ye_s_vel_histogram} is consistent: \texttt{Q7.chi0.8}
features on average shorter expansion timescales, corresponding to the
shift of the ejecta velocities to higher values. At the same time, it
features a much wider spread in the distribution of $\tau$. Inversely,
the $\tau$-distribution of \texttt{Q4.chi0.8} is more collimated and
features, in terms of bulk ejecta behavior (most populated bins), longer
timescales before and after the NSE corrections. Similarly
to~\cite{Radice2016}, the necessary adjustments to the expansion
timescales do not yield any statistically significant number of tracers
with $\tau$ smaller than $\approx 0.5~\rm{ms}$, making the neutron
freeze-out scenario less plausible for our models~\cite{Metzger2015}. The
predictions of Fig.~2 in Ref.~\cite{Schnabel2026} (as well as in Fig.~B2
in the Appendix therein.) suggest that some non-negligible free-neutron
abundance ought to be present in our models, when evaluated with our
reasonably optimistic bulk-ejecta parameters ($\rho_{{\rm ub}, 0}\approx
10^{7}~\rm{g}/\rm{cm}^{3}$, $s\approx 20-30\,k_{\rm B}/\rm{baryon}$, $v/c
\approx 0.3$, $Y_{e}\approx 0.05$, $T_{0}\approx 1\,\rm{GK}$). Since this
topic goes beyond the scope of our investigation, however, we will from
now on focus exclusively on the heavy element abundances.

Once the ejecta properties $(\rho_{0}, s, Y_{e})$, together with the
chosen initial temperature $T_{0}=6~\,\rm{GK}$, the expansion timescale
$\tau$, and the modular components -- strong, weak, and inverse nuclear
rates -- of the reaction network have been prescribed, \texttt{SkyNet}
has all the necessary information to start its computations. Reaction
calculations extend to $t = t_{\rm fin} = 10^{9}\,\rm{s}\approx
32\,\rm{years}$ in all cases, similarly to the choice made in
Ref.~\cite{Fernandez2017}. For such long times, the final abundances
given in terms of the mass number $A$ ought to be relatively stable and
hence roughly representative of the solar abundances which one may
compare against. The only exception are the heaviest nuclei $A>210$,
where alpha decay and fission still play a non-negligible
role. Fortunately, their importance in dictating the overall level of
variance in the final abundances across our simulations is minor -- with
solar data poorly constrained for such massive nuclei\footnote{To a
lesser extent, observational challenges also concern the $3$-rd
$r$-process peak~\cite{Puls2025}.}.

\subsection{$r$-process nucleosynthesis yields} 
\label{sec:r-process_yields}

Before presenting the results of the $r$-process nucleosynthesis obtained
using the tracer collection constructed in the previous section and the
nuclear-reaction network~\texttt{SkyNet}, we note that previous work on
the $r$-process nucleosynthesis from relativistic simulations of BHNS has
separated and identified the contributions from dynamical ejecta, winds
and the secular outflow~\cite{Just2015, Fernandez2017, Roberts2017,
  Desai2019, Fernandez2020}, employing, in some cases, also a treatment
for the evolution of the electron fraction when considering the radiative
transport of neutrinos. 
While neutrino transport is neglected here, this simplified
treatment allows us to isolate, within this approximation, the dependence
of the $r$-process abundances and EM counterparts on the dynamical ejecta
mass, geometry, entropy, and velocity, and hence to examine their
underlying dependence on $Q$ and $\chi_{_{\rm BH}}$.

\begin{figure*}
  \centering
  \includegraphics[width=0.66\textwidth]{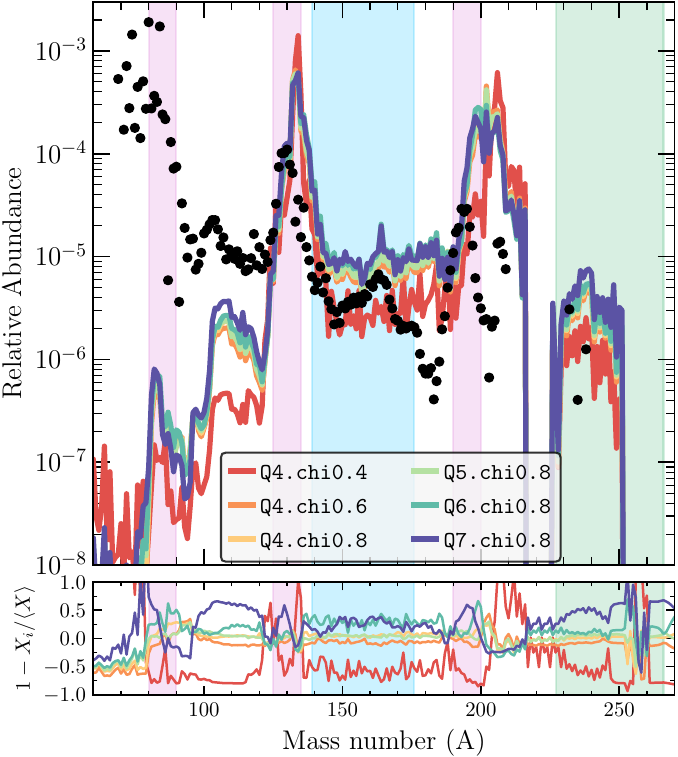}
  \caption{Relative abundances obtained as a result of $r$-process
    nucleosynthesis occurring in the dynamical ejecta and subsequent
    radioactive decay. The first, second and third $r$-process peaks are
    indicated with pink regions. Blue-, and green-shaded regions refer to
    the lanthanides and actinides, respectively. The solar $r$-process
    data is marked with black circles. The neutron-richness of the
    material leads to an order of magnitude greater abundances in the
    second and third $r$-process peak, and underestimation of the first
    solar $r$-process peak by multiple orders of magnitude in all
    cases. Note that while the data shows little variance close to the
    second and third peaks, differences appear between the second and
    third peak ($140 < A < 190$), in the actinide ($A \gtrsim 230$) and
    below the second peak
    region. Figure~\ref{fig:rprocess_abundances_cutoff} will show how
    this variance is related to a dependence on the entropy and velocity
    of ejecta. 
      }
  \label{fig:rprocess_abundances}
\end{figure*}

The collective results of the~\texttt{SkyNet} computations for the
various binaries considered are shown in
Fig.~\ref{fig:rprocess_abundances}, where the mass number $A$ is used on
the horizontal axis, while the relative abundances shown on the vertical
axis have been re-scaled with respect to the $A=132$ isotopes. Since this
rescaling makes the original relative abundances no longer sum up to
unity, the respective $A$-dependent mass fractions were re-normalized to
sum up to $1$ in order to facilitate comparison between the datasets. The
lanthanide group corresponds to the range of mass numbers $139 \leq A
\leq 176$ (light-blue shaded area). The other group of heavy elements,
also characteristic of strong $r$-process nucleosynthesis, is formed by
actinides and corresponds to $227 \leq A \leq 266$ (shaded
green). Approximate locations of the main $r$-process abundance peaks are
given by $A\approx 80-90$, $A\approx 130$ and $A\approx 195$ and are
shown as purple-shaded regions. Finally, the abundances of elements in
the solar system are marked with black filled circles. This data combines
the solar abundances measured in meteoritic data studies, and theoretical
calculations of Ref.~\cite{Lodders2009}, which we correct to include
solely the contribution from the $r$-process~\cite{Prantzos2019} (Table
$3$, last column), thus neglecting the $s$-process origin. The same
normalisation procedure with respect to the $A=132$ abundance is employed
also for the solar data.

A number of considerations can be made when examining in detail the
results shown in Fig.~\ref{fig:rprocess_abundances}. First, in none of
the cases does the $r$-process abundance pattern reproduce faithfully the
three-peak structure seen in the solar abundance data. Namely, the second
and third $r$-process peaks are an order of magnitude more abundant than
for the referential solar $r$-process, while the abundances near the
first peak are severely depleted, by many orders of magnitude. We
interpret this, consistently with prior literature~\cite{Korobkin2012}
(\cf Fig.$8$ therein), as the classic signature of very neutron-rich
dynamical ejecta. Namely, the material piles up at the heaviest peaks
via fission cycling and strong $r$-processing. These findings highlights
that the abundances we observe are exclusively the results of the
$r$-process nucleosynthesis occurring in the dynamical ejecta, while for
a complete picture also the secular outflows are necessary.

Secondly, the inset below the Fig.~\ref{fig:rprocess_abundances}, where
the relative difference between the mean abundance across simulations
$\langle X\rangle$ and their respective measurements is shown, reveals
that substantial differences appear in the data. Those differences are
mostly suppressed for all simulations at the locations of the second and
third $r$-process peak, but otherwise reach up to $50\%$ between
$r$-process peaks.

Third, the observed variations apply also to the lighter $r$-process
elements between the first and second $r$-process peaks, $A\approx
90-125$. Overall, the abundances up to the second $r$-process peak
$A\approx 125$ are rather similar irrespective of the mass ratio and BH
spin and exhibit a deficit of about one order of magnitude from the
solar-abundance data. The only marked outlier is the \texttt{Q4.chi0.4},
which is located decidedly below all the remaining simulations in that
region.

Fourth, beyond the second $r$-process peak, in the actinide region,
both the strongly and weakly tidal-disruptive cases (along with the
\texttt{Q4.chi0.4} configuration) remain closer to one another but
overproduce the elements when compared with the solar data.

Fifth, near the location of the third $r$-process peak, $A\approx 195$,
all configurations to some degree converge, yet their maxima remain
shifted by about $A \approx 10$ relative to the solar abundance pattern,
with an overall overproduction of heavy nuclei in that region by about an
order of magnitude.

Finally, the $r$-process nucleosynthesis at very low electron fractions
($Y_e \approx 0.05$) in our data generates transuranic elements and can
trigger fission cycling. Consequently, the dynamical ejecta from BHNS
mergers is expected to contain numerous heavy nuclides that undergo
fission and/or $\alpha$-decay. This is indeed confirmed by our data. More
specifically, beyond the third r-process peak -- where the heaviest
$r$-process products ($A \geq 220$) appear and solar data is relatively
poorly constrained -- the simulations exhibit a rather ordered trend: the
most strongly tidally disrupting mergers show the lowest relative
abundances, while the least disruptive mergers show the highest (this is
especially visible in the inset below the main figure). The
\texttt{Q4.chi0.4} simulation is, again, an outlier, and lies
systematically below all the other simulations. The systematic
dependence, largely governed by the $Q$ values, is likely a consequence
of the prolonged ejecta expansion timescales, an effect that we will
examine in more detail in
Appendix~\ref{subsec:rprocess_nucleosynthesis_entropy_vel_dep}.

We note that while our results share some similarities with the ones of
Refs.~\cite{Just2015, Fernandez2017, Roberts2017, Desai2019,
  Fernandez2020}, they are perceivably quite different. In particular,
our ensemble of simulations does not reproduce a three-peak structure,
but only the two heavier ones. At the same time, their expected locations
are systematically shifted toward higher mass numbers though overall
consistent across different configurations. Moreover, we observe an
overshoot with respect to the $r$-process solar abundances past the
lanthanide region, which appears to be mitigated in the results from the
literature when more unbound mass components, or neutrino irradiation are
considered.
In summary, notwithstanding the limitations of our nucleosynthetic
calculations and the corresponding yields, \ie the absence of neutrino
irradiation and the inclusion of dynamical ejecta only (which is,
however, expected to dominate over the secular ejecta for weakly
disruptive mergers in realistic mass-ratio
regimes~\cite{Fernandez2020,Duez2024a}\footnote{In their
$\mathcal{O}(1\rm{s})$-long simulations of post-merger $Q=4,6$,
$\chi_{_{\rm BH}}=0.75$ configurations,
Refs.~\cite{Hayashi2023,Hayashi2024} find a roughly equal split in the
total ejected mass between the dynamical and secular components.}), our
calculations indicate that the very neutron-rich dynamical ejecta from
BHNS mergers robustly produce heavy $r$-process nuclei, while not by
themselves reproducing the full abundance pattern. BHNS mergers may
nevertheless provide important contributions over parts of the mass
spectrum (see, \eg~\cite{Wanajo2022}). We note that calculations
including either a grey M1 neutrino treatment together with the secular
ejecta~\cite{Hayashi2023}, or Monte Carlo neutrino transport for the
dynamical ejecta~\cite{Martineau2024}, find approximately $70\%$ of the
ejecta mass at $Y_e\leq0.1$ and $30\%$ at $Y_e>0.1$.

\section{radiative-transfer and kilonova signals}
\label{sec:radiative_transfer_kilonova}

As a concluding aspect of our study of properties of the ejected material
from BHNS binaries, we next consider the corresponding multi-wavelength
EM signatures, which evolve over a broad range of timescales
and are powered by the radioactive decay of heavy $r$-process nuclei, \ie
the KN signal. To this scope, we employ the radiative-transfer
code~\texttt{POSSIS}~\cite{Bulla2019, Bulla2023}, whose main features we
briefly summarise below.

\subsection{Numerical setup}

\texttt{POSSIS} is a time-dependent, three-dimensional Monte-Carlo
radiative-transfer code designed to compute spectra, light-curves, and
polarization for arbitrary geometries in supernovae and kilonovae models.
In lieu of solving the equations of radiative transfer directly,
\texttt{POSSIS} requires opacities as input. Throughout the simulation,
the nuclear heating rates, thermalization efficiencies, as well as the
wavelength-dependent opacities all depend on local properties of the
ejecta and vary in time. In essence, the code simulates the propagation
of Monte-Carlo photon packets in a homologously expanding medium,
accounting for their interactions with matter through electron
scattering, as well as bound--bound transitions. The nuclear heating rates use the library of
Ref.~\cite{Rosswog2022} with an improved fitting formula which is aware
of the local electron fraction and ejecta velocity. Heating rates therein
are based on the output of the \texttt{WinNet} code and take into account
electron-fraction values as low as $Y_{e}=0.05$ (which is also the lower
limit of our simulations, see Tab.~\ref{tab:dyn_ejecta_properties}). In the more recent version of \texttt{POSSIS}, an improved, time-
and density-dependent treatment of thermalization efficiencies of
Ref.~\cite{Wollaeger2017} is employed. Finally, the opacities are
informed by the tabulated data of Ref.~\cite{Tanaka2020_opacity}. We note
that because the low electron fraction of our ejecta actually corresponds
to values below the minimum of the table therein, the opacities are obtained
assuming uniform value of $Y_{e}=0.1$ in the ejecta, which corresponds
still to a very neutron-rich regime; stated differently, existing tables
limit the ability to track spatial variations of $Y_e$ for
$Y_{e}<0.1$. Thanks to its versatility, \texttt{POSSIS} has been used in
 variety of applications, including the studies of KN observables
associated with compact binary mergers~\cite{Bulla2018, Bulla2019,
  Dietrich2020, Anand2021, PerezGarcia2022, Neuweiler2022,Mathias2023,
  Markin2025, Neuweiler2025b}, EM signatures of tidal disruption
events~\cite{Charalampopoulos2022, Leloudas2022}, as well as polarization
characteristics of supernovae~\cite{Inserra2016}.


\subsection{Coupling \texttt{FIL} and \texttt{POSSIS}}


An important prerequisite for using the ejecta in our
radiative-transfer calculations is the assumption of homologous expansion,
\ie $x_i=v_i t$ with approximately constant asymptotic velocities.
At our extraction time, $t-t_{\rm mer}\approx15\,{\rm ms}$, exact
homology is not expected to be fully established. Indeed,
Ref.~\cite{Neuweiler2022} found a dependence of synthetic light curves on
the extraction time, with differences generally $\lesssim1\,{\rm mag}$ for
their early-time extractions and decreasing to $\lesssim0.4\,{\rm mag}$ for
$t-t_{\rm mer}\gtrsim80\,{\rm ms}$.

Nevertheless, the bulk dynamical ejecta in our simulations is already
predominantly radial at the extraction time. As discussed in
Sec.~\ref{sec:geometric_XY_XZ}, the non-radial velocity contribution is
$\lesssim5\%$ in the densest equatorial regions, remains below
approximately $10\%$ away from the orbital plane, and is about $2\%$ in
the more tenuous ejecta. Moreover, comparable extraction times of
$\mathcal{O}(10\,{\rm ms})$ have been employed in recent NR-informed
radiative-transfer calculations~\cite{Markin2026,Neuweiler2025b}.
We therefore regard the homologous rescaling as a source of systematic
uncertainty in the absolute light curves and focus primarily on relative
trends among models treated with the same prescription.

\begin{figure}
  \centering
\includegraphics[width=0.99\columnwidth]{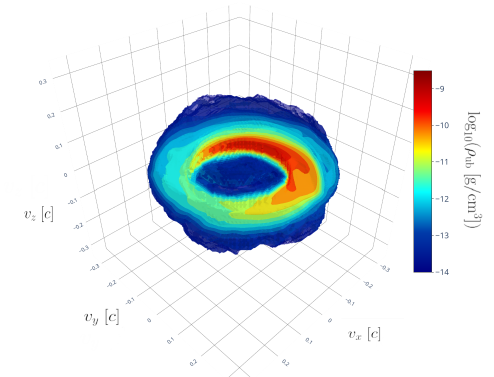}
  \caption{Volume rendering of the rest-mass density for the
    \texttt{Q6.chi0.8} binary interpolated onto a
    grid of uniform resolution and rescaled to represent the ejecta at $t
    = 0.1\,\mathrm{d}$ after the merger under the assumption of
    homologous expansion. Under this assumption, the velocity components
    may be used to label the axes, with most of the ejecta within $\vert
    v \vert \leq 0.3\,c$ and the densest, red region $\vert v \vert
    \approx 0.25\,c$. After rescaling, the grid spans approximately $1.5
    \times 10^{14}\,\mathrm{cm}$ in each direction, with maximum ejecta
    densities of $\rho_{\mathrm{ub,max}} \approx 10^{-9} \, \mathrm{g /
      cm^{3}}$.}
  \label{fig:ejecta_input_possis}
\end{figure}

Our ejecta data are interpolated onto a uniform \(100^3\) grid and
rescaled to \(T=0.1\,{\rm d}\) assuming homologous expansion; an example
is shown in Fig.~\ref{fig:ejecta_input_possis}\footnote{For an alternative approach to initializing
radiative transfer calculations, see \eg~\cite{Collins2022} where tracer
particles in a simulation are used to inform the initial state by
utilizing their final measured velocities.}.

The rather anisotropic distribution of matter shown in the figure
anticipates the fact that the synthetic observables from \texttt{POSSIS}
will display considerable dependence on observer's location. To probe the
full dependence, we extend the standard setup in the code to incorporate
azimuthal variation in the viewing angle. More concretely, we discretize
the polar angle $\theta$ into $11$ values $\theta_{i}$, with
$\cos\theta_i \in [0,1]$ with an increment of $\Delta \cos\theta :=
\cos\theta_{i+1} - \cos\theta_{i} = 0.1$. This setup then provides both a
polar view with $\theta=0^{\circ}$ (\textit{face-on} view), an equatorial
view with $\theta=90^{\circ}$ (\textit{edge-on} view), as well as the
intermediate angles. On the other hand, to capture the anisotropy in the
$\varphi$ direction, we use increments of $\Delta \varphi = 30^{\circ}$
to provide a full and comprehensive azimuthal coverage. As a result, we
considered a total of $N_{\rm obs} = (N_{\theta}-1) \times N_{\varphi} + 1 = 10
\times 12 +1 = 121$ independent viewing angles (since at $\theta=0$ the dependence
on $\varphi$ drops out). Finally, we set the number
of photon packets in our simulations to $N_{\rm ph}=10^{6}$ and carry out
the computations until $T=60\,\rm{d}$.

\subsection{Bolometric luminosities}
\label{sec:Lbol_and_angles}

\begin{figure*}
  \centering
  \includegraphics[width=0.66\textwidth]{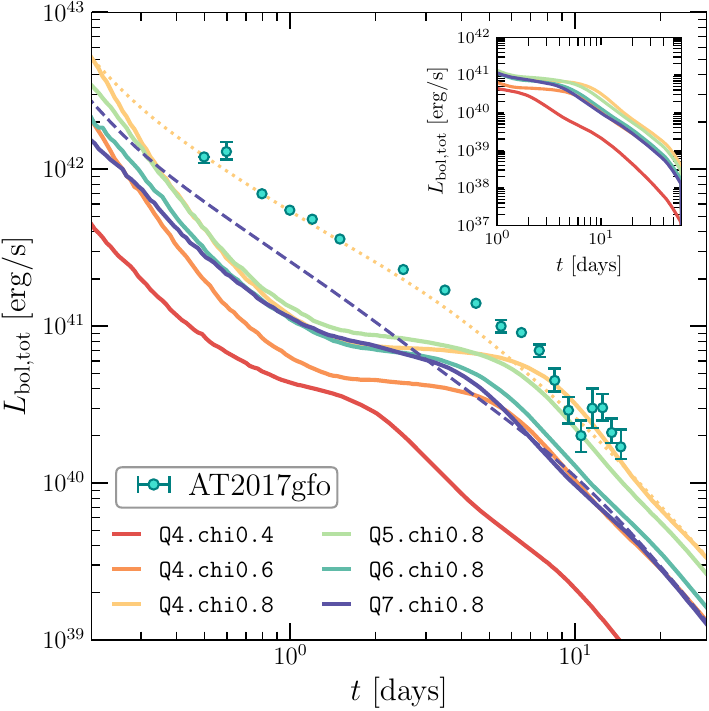}
  \caption{Total bolometric luminosities for the different binaries
    considered when scaled to a distance of $40\,\rm{Mpc}$. The magnitude
    of the EM emission in strong disruption cases \texttt{Q4.chi0.8} and
    \texttt{Q5.chi0.8} is broadly consistent at $t\gtrsim 10\,\rm{d}$
    with the evolution of the late-time lanthanide-rich kilonova
    component of AT2017gfo whose bolometric luminosity follows from
    Ref.~\cite{Waxman2017}. Heating rates associated with the radioactive
    decay and responsible for long-lived emission are shown for
    \texttt{Q4.chi0.8} and \texttt{Q7.chi0.8} in terms of dotted-yellow
    and dashed-blue lines, respectively.}
  \label{fig:possis_total_lbol}
\end{figure*}

We start with Fig.~\ref{fig:possis_total_lbol} to present the results of
the radiative-transfer simulations performed with~\texttt{POSSIS} and
employing the dynamical ejecta from BHNS simulations described
earlier. More specifically, we report in the figure the (total)
bolometric luminosities $L_{\rm bol, tol}$ focusing at times past $t -
t_{\rm mer} = 0.2\,\rm{d}$\footnote{Considering the very large timescales
involved in the EM emission, for which $t - t_{\rm mer} \sim t$, we will
omit the merger time $t_{\rm mer}$ hereafter.} and over a time interval
of $30\,\rm{d}$ ($60\,\rm{d}$ in the inset panel), where different
colours are used to refer to the different BHNS binaries considered (the
same colour convention will be adopted also in the figures that follow),
while the filled circles represent the bolometric luminosity data
reported for AT2017gfo in~\cite{Waxman2017}.

Although the luminosities generally remain at the level of
$10^{42},\mathrm{erg,s^{-1}}$ at $t = 0.2\,\mathrm{d}$ and decline
steadily to about $10^{40},\mathrm{erg,s^{-1}}$ by $t \approx
10\,\mathrm{d}$ after the merger, clear differences appear in the
bolometric light-curves, both in the magnitude and in the short-term
behavior. For instance, the substantially smaller dynamically ejected
mass in \texttt{Q4.chi0.4} leads to a bolometric luminosity that is
roughly one order of magnitude smaller than in all other
binaries. Likewise, the early-time emission ($t < 5\,\mathrm{d}$) of the
binary \texttt{Q4.chi0.6} differs from all the others by a factor of
several. At the same time, the behavior of \texttt{Q4.chi0.8} and
\texttt{Q5.chi0.8} at $t\geq 8\,\mathrm{d}$ suggests that the
lanthanide-rich ejecta of those simulations reproduce the late-time
behavior of the GW170817 kilonova even though its origin was a BNS
rather than a BHNS system. We note for completeness that this late-time
behavior is also influenced by effects such as photon reprocessing and
thus not solely determined by material's composition.

The light-curves in Fig.~\ref{fig:possis_total_lbol} also reveal that the
influence of increased BH spin at fixed mass ratio is particularly
important for early and intermediate times. Higher spins allow the
production of larger amounts of dynamical ejecta. We observe that the
greater ejecta mass increases radioactive heating, thereby boosting the
early- and intermediate-time luminosity. In the strong tidal disruption
cases -- \eg the binaries \texttt{Q4.chi0.8} and \texttt{Q5.chi0.8} --
the light-curves display a shallow decay at $\sim 10^{41}\,\rm{erg/s}$
around $t \approx 1$-$3\,\rm{d}$. This phase arises in tandem with a
steady decrease in the deposition curve, which describes the rate at
which radioactive decay energy is thermalized in the ejecta, and is a
common finding in radiative-transfer simulations and in actual
observations, amplified at high ejecta masses. The onset of the major
luminosity decline after the shallow decay is slightly delayed in more
disruptive mergers. For instance, at the level $L_{\rm bol,tot}\sim
10^{41}\,\rm{erg/s}$, \texttt{Q4.chi0.8} begins to fade around $t \approx
7\,\rm{d}$, whereas in the less massive disruption \texttt{Q7.chi0.8} the
decline occurs already by $t \approx 3\,\rm{d}$. This difference is
because more massive ejecta have longer diffusion times, storing the heat
for a longer period and sustaining the luminosity before the radioactive
power eventually falls below the photon escape rate.

Despite the variance discussed for the early- and intermediate-time
luminosities, at late times, \ie for $t\geq 8-10\,\mathrm{d}$ all the
binaries exhibit rather identical light-curves, and are only shifted in
magnitude with respect to one another (see inset). This is because, at
late times, the ejecta expansion reaches an optically thin regime, where
the bolometric luminosity follows a simple energy-deposition curve,
plotted as dotted yellow and dashed blue lines for the \texttt{Q4.chi0.8}
and \texttt{Q7.chi0.8} binaries, respectively. Assuming that the heating
rates and the thermalization efficiencies in each case are comparable
(which is a reasonable assumption in case of our simulations) the
magnitude of $L_{\rm bol, tot}$ at late times is then governed primarily
by dynamical ejecta mass. In other words, while for earlier emission
stages the geometry of the ejecta also plays a role, for the late time
(optically thin, isotropic) emission is mostly governed by the amount of
ejected mass.

During the last stage of the bolometric EM emission, the mass ratio at
fixed BH spin is a decisive factor in the magnitude of luminosity,
establishing the ordering of the simulations at these epochs. For
example, at $t = 10\,\rm{d}$, the luminosity of the \texttt{Q7.chi0.8}
($M_{\rm ej}\approx 0.03\,M_{\odot}$) simulation is roughly a factor of
five lower than that of \texttt{Q4.chi0.8} ($M_{\rm ej}\approx
0.06\,M_{\odot}$); these effects are clearly correlated with the amount
of dynamical ejecta present. Indeed, the \texttt{Q4.chi0.6} and
\texttt{Q7.chi0.8} simulations both feature $M_{\rm ej}\approx
0.03\,M_{\odot}$ and at $t>5\,\rm{d}$ are very close to each other, in
full agreement with the expectations from the optically-thin regime.
Although AT2017gfo was not produced by BHNS system, the comparison is
still reasonable given the character of the late-time emission and to
obtain a quantitative estimate of how much BNS and BHNS emissions can
differ in the cases considered here.

\subsection{Dependence on the viewing angle}

\begin{figure*}
  \centering
  \includegraphics[width=0.48\textwidth]{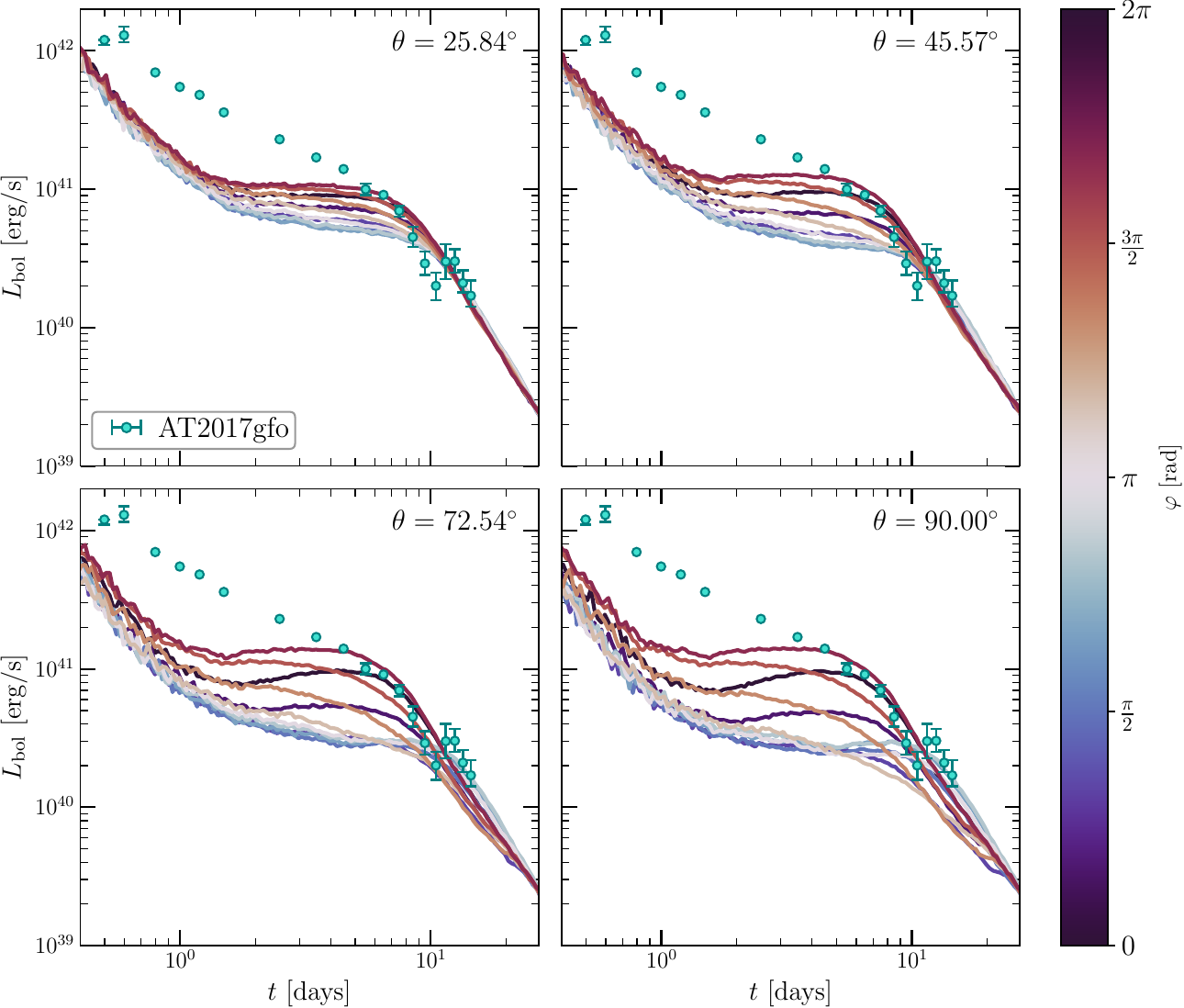}
  \hspace{0.5cm}
  \includegraphics[width=0.48\textwidth]{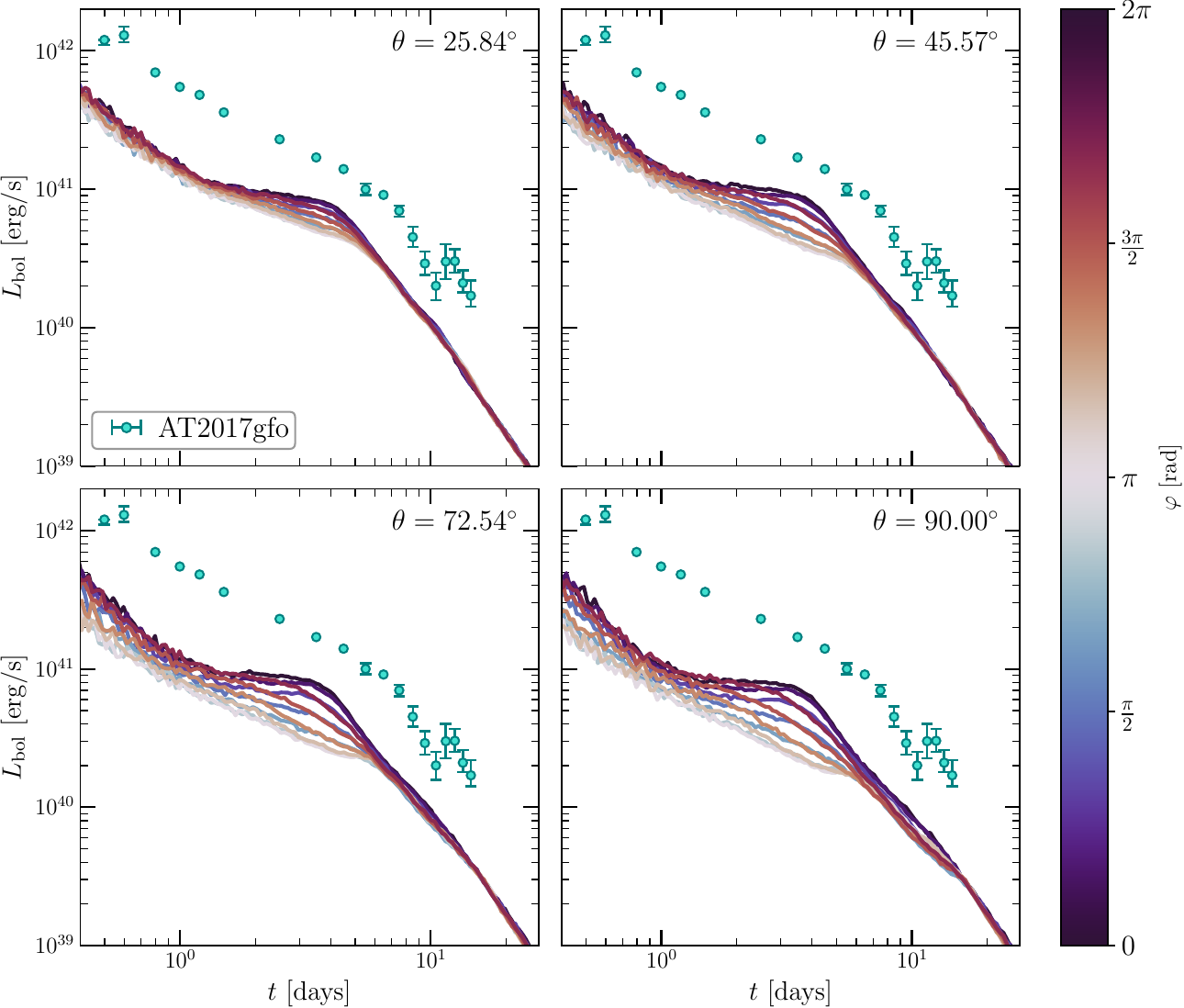}
  \caption{\textit{Left panels:} Dependence on the viewing angle of the
    evolution of the bolometric luminosity (see colormap) as computed
    from the strong-disruption binary \texttt{Q4.chi0.8}. The
    data from AT2017gfo is shown for a comparison with
    cyan filled circles~\cite{Waxman2017}. Note that only the late-time
    bolometric luminosity has a magnitude comparable with AT2017gfo data
    and that the dependence on the viewing angle is substantial, leading
    to angles where temporary shallow decay or even a re-brightening is
    possible. \textit{Right panels:} The same as on the left but for the
    weak-disruption binary \texttt{Q7.chi0.8}.}
  \label{fig:possis_lbol_angular_Q4.chi0.8}
\end{figure*}

As anticipated above, the anisotropic distribution of the ejecta
naturally leads to a dependence of the bolometric luminosities on the
viewing angle, which we will discuss next for two particularly
interesting cases of disruptive mergers: \texttt{Q4.chi0.8} (left panels
of Fig.~\ref{fig:possis_lbol_angular_Q4.chi0.8}) and \texttt{Q7.chi0.8}
(right panels of Fig.~\ref{fig:possis_lbol_angular_Q4.chi0.8}) as
representatives of strong and weak disruption binaries, respectively (we
recall that the ejected mass in the two binaries is $M_{\rm ej}\approx
0.06\,M_{\odot}$ and $M_{\rm ej}\approx 0.03\,M_{\odot}$). The various
panels in Fig.~\ref{fig:possis_lbol_angular_Q4.chi0.8} provide a
comprehensive and systematic overview of the viewing angle dependence in
the observed luminosities, for four representative polar angles:
$\theta\in \{25.84^{\circ}, 45.57^{\circ}, 72.54^{\circ},
90.00^{\circ}\}$ and $12$ equally-spaced azimuthal angles (where
naturally $360^{\circ} = 0^{\circ}$ and is hence already covered). All
the luminosities are similarly rescaled to $40\,\rm{Mpc}$ and refer to
times $t\geq 0.4\,\rm{d}$ for a more focused view. Additionally, they are
accompanied by the isotropically equivalent bolometric luminosity
reported for the AT2017gfo data~\cite{Waxman2017}, represented by
cyan-filled circles with visible error bars. The choice of a cyclic
colormap for the azimuthal $\varphi$ angle is representative of the
$2\pi$ periodicity, where angles close to $0^{\circ}$ (\eg $\pm
30^{\circ}$) are perceptually close.

A rapid inspection of Fig.~\ref{fig:possis_lbol_angular_Q4.chi0.8}
reveals that a clear dependence on the viewing angle is present both for
early ($t\lesssim 1\,\mathrm{d}$) and intermediate times ($3 \lesssim
t/\mathrm{d} \lesssim 10$). More specifically, for early times, the
light-curves near the polar direction (\ie for $\theta\approx
25.84^\circ$ and $45.57^\circ$) exhibit the highest luminosities,
exceeding the more equatorial views (\ie for $\theta\approx 72.54^\circ$)
by up to a factor of $\sim2$ at peak. As $\theta$ increases toward the
equator, the early-time luminosity systematically decreases, and the
spread among different azimuths becomes more pronounced. By
$\theta=90.00^\circ$, the curves are systematically dimmer at early
epochs, reflecting the larger optical depths encountered in equatorial
viewing directions, where the line of sight intersects the dense,
lanthanide-rich tidal ejecta.

The variations in luminosity with the azimuthal angle $\varphi$ at fixed
polar angle $\theta$ are generally modest at early times, particularly
for the more polar views. However, by $t\gtrsim 2\,\mathrm{d}$, azimuthal
modulations become more visible, especially at intermediate and
equatorial $\theta$. The azimuth at which the luminosity peaks tends to
align with directions corresponding to the geometric ``opening'' of the
tidal ejecta. This is in qualitative agreement with the findings of \eg
Ref.~\cite{Kawaguchi2024a}, which reports that the maximum equatorial
luminosity occurs when the line of sight is aligned with the center of
the ejecta's azimuthal extent, \ie viewing into the broadest projected
cross-section of the ejecta in the orbital plane (see the upper two
panels in Fig.~9 therein). In our dataset (left panel), this brightening
manifests as ``warm''-colour curves (orange-brown in the upper part of
the colormap, plus the black curve at the neighbouring
$\varphi=0^{\circ}$ angle) clustered near $\varphi$ values corresponding
to this opening, with visually ``cold'' curves associated with azimuths
that intersect the ejecta's trailing edges (and hence on the opposite
side). This underlying geometric effect can be verified also in
Fig.~\ref{fig:Q4.chi0.8_slice_2d_xy}, where the dynamical ejecta is
directed toward positive $x$ and even more strongly toward negative $y$,
corresponding to the luminosity-maximizing azimuthal angle of $\varphi
\approx {11\pi}/{6} \approx 330^{\circ}$.

At later times, \ie for $t\gtrsim 10\,\mathrm{d}$, the overall
$\theta$-dependence becomes weaker, as the ejecta expand and enter the
optically-thin regime, rendering the emission approximately
isotropic. Nonetheless, the polar views remain systematically brighter,
and the azimuthal structure is still visible for equatorial and
intermediate latitudes, hinting at the long-lasting effects of an
asymmetric distribution in the material. As one would expect for a
geometrically induced effect, the shallow decay and minor re-brightening
in a direction that is diametrically opposite ($\varphi\approx
{5\pi}/{6}$) to that of the maximal luminosity ($\varphi={11\pi}/{6}$) is
achieved much later and at $t\approx 10\,\rm{d}$ (see the light blue-gray
curve).

Interestingly, for the most favorable azimuthal orientations at $\theta
\approx 90^\circ$, the luminosity can temporarily exceed that of the
corresponding (more) polar views at intermediate times ($t \sim
1$-$3\,\mathrm{d}$). This inversion of the polar-equatorial hierarchy
occurs when the line-of-sight is aligned with the ejecta's widest opening
in the orbital plane, reducing the effective optical depth along the
equator and allowing a more rapid photon escape. In these cases, the
light-curves also display a shallower decline, and in some instances even
a short-lived plateau or mild re-brightening phase between
$t\sim1{-}3\,\mathrm{d}$, before resuming the overall decay trend. Such
behavior suggests that geometric asymmetries and diffusion effects
dependent on the viewing angle can temporarily compensate the increased
opacity associated with equatorial lines-of-sight (see also
Ref.~\cite{Kawaguchi2024a}).

We note that in contrast to the strongly disruptive \texttt{Q4.chi0.8}
case, the spread between different viewing angles is markedly reduced for
the weak-disruption event \texttt{Q7.chi0.8}, both in terms of polar and
azimuthal variations (see right panels in
Fig.~\ref{fig:possis_lbol_angular_Q4.chi0.8}). At early times ($t\lesssim
1\,\mathrm{d}$), polar views remain slightly brighter, but the difference
relative to equatorial orientations rarely exceeds a factor of $\sim
0.2{-}0.3$. This angular invariance reflects the more uniform ejecta
geometry and smaller overall mass, which reduce both the absolute optical
depths and contrast between the lines of sight. The azimuthal modulation
at fixed $\theta$ is similarly weak, with the brightest and dimmest
orientations differing only marginally; in particular, no strong
enhancement is observed for azimuths corresponding to the geometric
opening of the tidal ejecta, as was the case for the binary
\texttt{Q4.chi0.8}.

Another notable difference between the two strong/weak disruptions is the
significantly shorter (or even nonexistent) shallow decay or temporary
re-brightening phase in the equatorial light-curves, save for the most
advantageous angles. Indeed, across all values of $\theta$ and $\varphi$
considered, the luminosities predominantly follow a smooth and monotonic
decline from peak to late times, consistent with the faster diffusion
times expected for the smaller ejecta mass and reduced geometric
asymmetry (\cf the much more uniform ejecta profile in the equatorial
plane illustrated in Fig.~\ref{fig:Q7.chi0.8_slice_2d_xy}). By $t\gtrsim
5\,\mathrm{d}$, the curves for all polar angles nearly converge,
indicating that the system has entered a regime where the angular
dependence of the optical depth is minimal and the emission is
effectively quasi-isotropic.

As a concluding remark, it is important to note that the synthetic
light-curves for \texttt{Q7.chi0.8} lie systematically below the
AT2017gfo measurements across most epochs and viewing angles, with the
largest deficits occurring at $t\sim 1-3\,\mathrm{d}$. Because this
offset reflects the reduced ejecta mass -- and hence reduced radioactive
heating -- it is natural to conclude that BHNS binaries with higher mass
ratios, and thus with weak disruptions, are expected to provide, for any
given distance, the weakest peak luminosities and the most rapid
declines. On the other hand, the strong disruption case
\texttt{Q4.chi0.8} matches the bolometric luminosities of AT2017gfo after
$t \gtrsim 6\,\mathrm{d}$ due to the larger dynamical ejecta mass
involved.

\begin{figure*}
  \centering
  \includegraphics[width=1.0\textwidth]{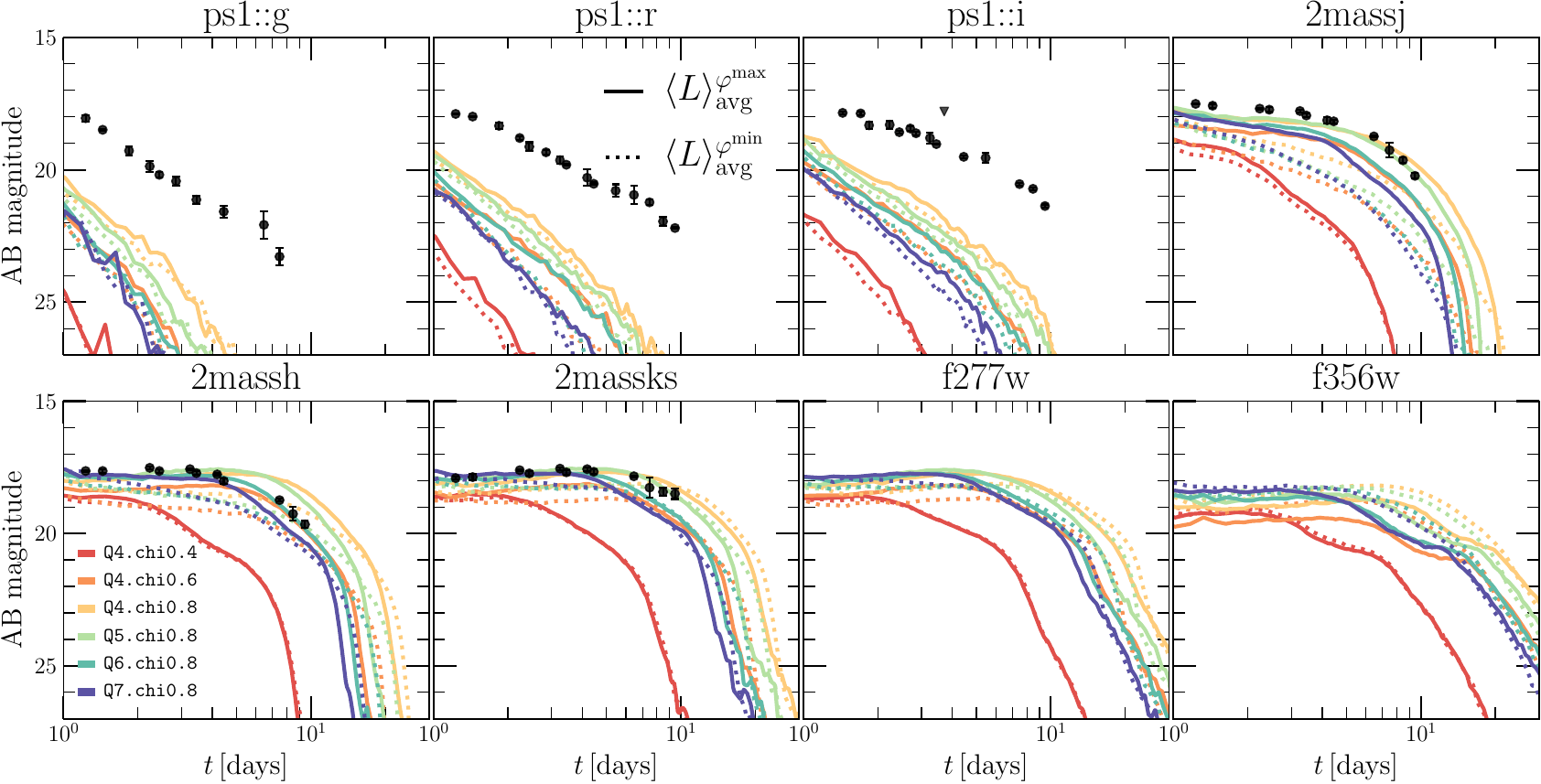}
  \caption{Photometric luminosities given in terms of the apparent
    magnitude, computed from the simulation data for the various binaries
    and compared with AT2017gfo measurements, utilizing the compiled data
    of Ref.~\cite{Singh2026}. From the top-left to the bottom-right
    corner, different panels refer to the various filters from
    Pan-STARRS1 collaboration (optical $\rm{g}$, $\rm{r}$, $\rm{i}$
    bands; first triplet), 2MASS (near-infrared $\rm{J}$, $\rm{H}$,
    $\rm{Ks}$ bands; second triplet), and JWST NIRCam (f277w, f356w).
    The response functions of those filters move incrementally toward NIR
    wavelengths. For all the light-curves presented, the polar viewing
    angle corresponds to a latitude of $\theta=25.84^{\circ}$. For each
    panel, different line types refer to two azimuthal averages: $\langle
    L\rangle_{\rm{avg}}^{\varphi^{\rm max}}$ for angles facing the ejecta
    (solid line) and $\langle L \rangle_{\rm{avg}}^{\varphi^{\rm min}}$
    (dotted line) for the opposite direction.}
  \label{fig:possis_photometric_luminosities_diff_filters}
\end{figure*}

\subsection{Photometric luminosities: comparison with AT2017gfo}
\label{sec:Lpho_and_angles}

To complement the EM picture constructed above in terms of the bolometric
luminosities, we now turn to comparing specific band-pass properties and
related evolution with the AT2017gfo observations. To that end, in
Fig.~\ref{fig:possis_photometric_luminosities_diff_filters} we report the
apparent magnitude predicted for our simulation data as measured in $8$
different bands, covering optical to near-infrared wavelengths, as
implemented via \texttt{SNCosmo}~\cite{sncosmo}, and rescaled to a
distance of $40\,\rm{Mpc}$ corresponding to AT2017gfo~\cite{Cantiello2018}.
The two final filters inspected here correspond
to wide, long-wavelength channels of JWST's near-infrared camera
(NIRCam), namely f277w and f356w\footnote{Note that the tabulated
opacities of Ref.~\cite{Tanaka2020_opacity} implemented in
\texttt{POSSIS} encode the wavelength-dependence up to
$\lambda=3.5\mu\mathrm{m}$ and thus correspond only partially to the
wide-band extent of f356w. For the same reason, our presentation does not
go beyond this filter.}. Furthermore, a viewing angle of $\theta =
25.84^{\circ}$ is chosen in all cases, which is compatible with the
inferred one for GW170817 based on the observed superluminal motion of
the jet~\cite{Mooley2022}. Finally, time-and band-dependent measurements
of AT2017gfo compiled in Ref.~\cite{Singh2026} (and references therein) are plotted on each
subplot as markers, with inverted triangles indicating an
upper limit.

Note that, for each binary configuration, two different light-curves are
shown and report respectively $\langle L \rangle_{\rm{avg}}^{\varphi^{\rm
    max}}$ (solid lines) and $\langle L \rangle_{\rm{avg}}^{\varphi^{\rm
    min}}$ (dotted lines), which denote azimuthal averages over
neighbouring angles of the largest or the smallest cumulative bolometric
luminosity, respectively. In other words, once a value for $\theta$ is
chosen among those reported in
Fig.~\ref{fig:possis_lbol_angular_Q4.chi0.8}, $\varphi^{\rm max}$
($\varphi^{\rm min}$) corresponds to the azimuthal viewing angle
$\varphi$ which maximizes (minimises) the radiated bolometric energy
\begin{equation}
  \label{eq:Ebol}
  E_{\rm bol}(\theta, \varphi) := \int_{t =
    0\,\rm{d}}^{t=30\,\rm{d}} L_{\rm bol}(t, \theta,\varphi) dt\,.
\end{equation}
Subsequently, the photometric luminosity curve is averaged over the
$\varphi^{\rm max}$ ($\varphi^{\rm min}$) angle and two of its
neighbouring angles (\ie $\pm 30^{\circ}$ in either direction) and the
corresponding averaged luminosities obtained in this way are denoted with
$\langle L \rangle_{\rm{avg}}^{\varphi^{\rm max}}$ and $\langle L
\rangle_{\rm{avg}}^{\varphi^{\rm min}}$, respectively.

What appears clearly from
Fig.~\ref{fig:possis_photometric_luminosities_diff_filters} is that the
magnitude of the emission in the near-infrared ($\rm{J}$, $\rm{H}$,
$\rm{Ks}$) bands is overall consistent with AT2017gfo throughout all the
epochs in case of strong tidal disruptions (\texttt{Q4.chi0.8},
\texttt{Q5.chi0.8}) due to the large dynamical ejecta component of
$M_{\rm eje}\approx 0.06M_{\odot}$. At the same time, the predicted
values fall below the observed magnitudes for weaker disruptions, and
thus, for more moderate and lower dynamical ejecta masses. Furthermore,
regardless of the values of $Q$, $\chi_{_{\rm BH}}$, and of the
(averaged) viewing angle, the luminosity increases for filters that are
progressively redder. More precisely, it is possible to observe a rapid
decrease of the luminosity in the optical bands ($\rm{g}$, $\rm{r}$, and
$\rm{i}$ bands, first three panels), represented by a fall-off of more
than $5$ orders of magnitude within $10$ days and thus with a slope
greater than $0.5\,\rm{mag}/\rm{day}$. The small magnitude at optical
wavelengths is related to low electron fraction in our ejecta, and hence
to the abundance of lanthanides with high opacities in the $\lambda
\approx 500-800\,\rm{nm}$ regime. Because the overall magnitude is
dictated by the amount of ejected matter, a clear correlation is present
with $Q$ and $\chi_{_{\rm BH}}$ values for any given filter. At the same
time, filters corresponding to the (near) infrared wavelengths (top-right
panel and the entire bottom row) -- represented by the $\rm{J}$, $\rm{H}$
and $\rm{Ks}$ band filters of 2MASS as well as the f277w and f356w
filters of JWST -- feature a much richer behavior, differing
significantly not only across the mass ratio and BH spin, but also
displaying prominent differences between oppositely placed viewing
angles. For such bands, all magnitudes at viewing angles aligned with the
ejecta are approximately constant at epochs $t<7\,\rm{d}$ and
interestingly, display a small increase for strong disruption
datasets. On the other hand, anti-aligned viewing angles (\ie averages
around $\varphi^{\rm min}$) tend to display, if not constant magnitude,
then a steady and small decline of $0.3-0.5\rm{mag}/\rm{day}$ in the
early phase and $1-1.5\rm{mag}/\rm{day}$ later, an indication of strong
inherent EM emission at those wavelengths. Indeed, for near infrared
(NIR) filters represented in the two panels in the bottom-left part of
the figure, strong-disruption cases \texttt{Q4.chi0.8} and
\texttt{Q5.chi0.8} actually have comparable magnitudes at $t\approx
10\,\rm{d}$ regardless of the viewing angle.

Overall, the evolution in the JWST filters provides important insight in
case of a future KN detection -- and for a progenitor consistent with a
BHNS interpretation. Namely, should a hypothetical KN observation only be
made at late times, or without complementary GW data to ascertain the
nature of a potential binary, the comparison shown in
Fig.~\ref{fig:possis_photometric_luminosities_diff_filters} suggests that
conclusions can only be drawn relative to the near-infrared
wavelengths. This fact also becomes clear when comparing our strongly
disruptive BHNS binaries with the AT2017gfo counterpart for the cases
where data is available. Moderate to late-time ($t\sim10-20\,\rm{d}$)
wide-infrared emission for strong disruptions is universally
characterized by $\langle L \rangle_{\rm{avg}}^{\varphi^{\rm min}} >
\langle L \rangle_{\rm{avg}}^{\varphi^{\rm max}}$ due a longer effective
optical depth in the high-opacity material and the delayed arrival of
photon packets (\cf f277w and f356w results for \texttt{Q4.chi0.8} and
\texttt{Q5.chi0.8} in the last two panels). Finally, and importantly, our
results underline that late-time KN emissions of BNS and BHNS might look
alike in the near- and medium-infrared bands. Therefore, the observation
and modelling of a KN signal at early times and in the UV-optical bands
is of utmost importance.

\subsection{Photometric luminosities: comparison with S190814bv}

\begin{figure*}
  \centering
  \includegraphics[width=0.9\textwidth]{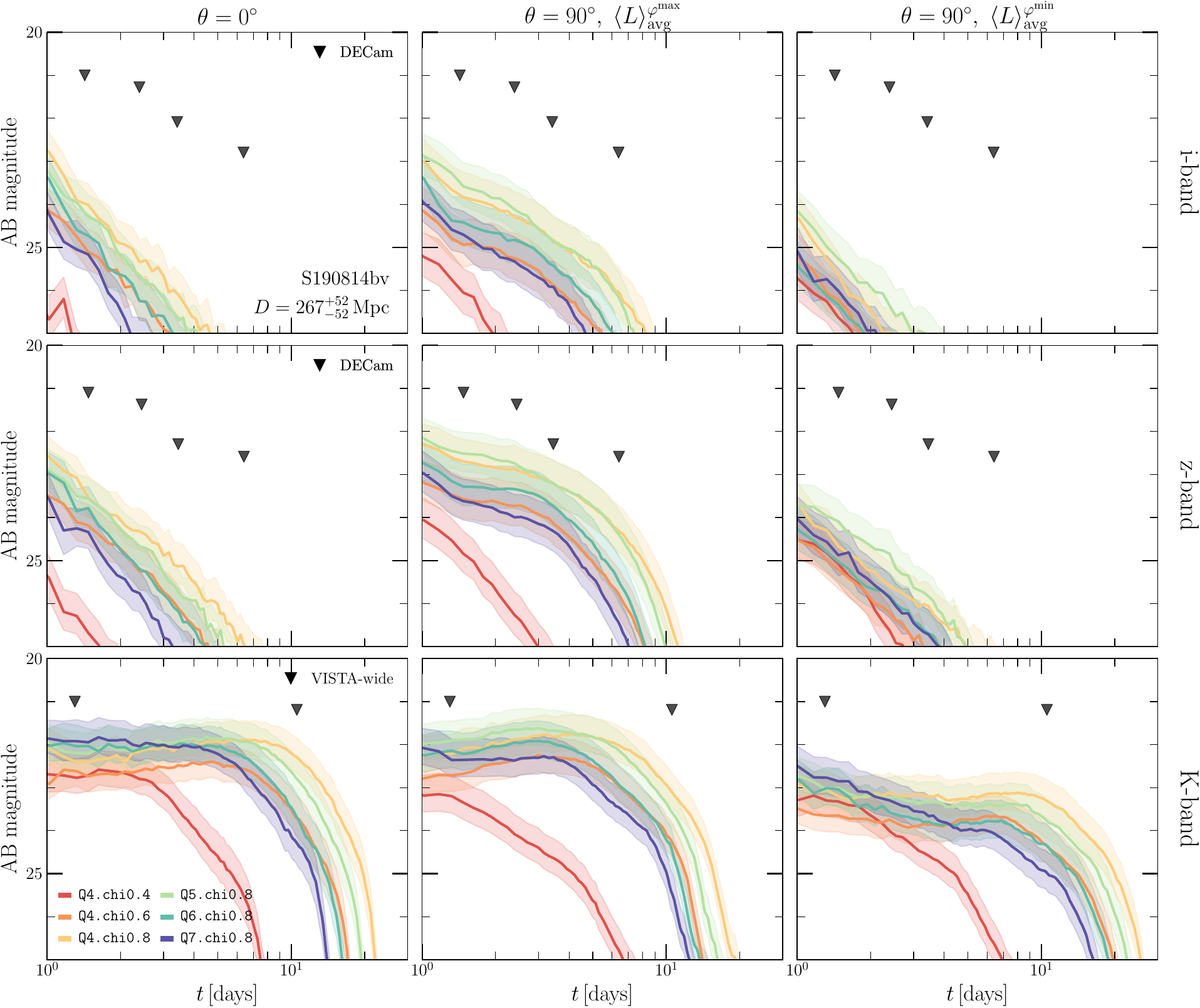}
	\caption{Comparison of the photometry for the event S190814bv~\cite{Ackley2020, Thakur2020} with the
    synthetic light-curves produced with the simulation data from
    different binaries. Three bands, ${\rm i}$, ${\rm z}$ and ${\rm K}$,
    are shown in consecutive rows and with an identification of the
    relevant instrument. In all cases, the first column represents a
    face-on (\ie independent of the azimuthal direction) viewing angle,
    with the second and third columns showing the edge-on case, with
    maximal and minimal azimuthal angle averages. The uncertainty bands
    on each light-curve are obtained by utilizing the uncertainty of the
    distance to S190814bv progenitor and are shown as transparent
    colour-shaded areas.}
  \label{fig:photometric_luminosities_event_comparison}
\end{figure*}

Next, to conclude our discussion on the photometric luminosities, we
consider a comparison between the magnitudes estimated from
\texttt{POSSIS}, and on the basis of our simulations, and the
non-detection bounds on the EM counterpart following a search triggered
by a GW detection. While none of our binary configurations are tailored
to the inferred binary parameters, it is nevertheless interesting to ask
whether they are compatible with the bounds set on the some of the
potential EM counterparts~\cite{Andreoni2020, Abbott2020, Anand2021,
  Abbott2021, LIGOScientific2024, Pillas2025, Ahumada2025}. Among the
established BHNS EM candidates with known upper bounds at hand, we opt
here to consider the event GW190814~\cite{Andreoni2020}, which is also
consistent with a more likely BBH progenitor~\cite{Nathanail2021}. Our
choice is motivated by the fact that in the case of S190814bv, both the
temporal range and the availability of upper-limits data in different
filters~\cite{Ackley2020, Thakur2020} matches the regime covered by our
simulations. Additionally, the distance of $267\pm{52}\,\rm{Mpc}$ inferred for the event~\cite{LIGOScientific_2019GCN} is
representative for putative future KN detections and currently 
observed KNe (see Ref.~\cite{Singh2026} for details).

To this end, in Fig.~\ref{fig:photometric_luminosities_event_comparison}
we present a comparison of our synthetic light-curves in terms of the AB
magnitude with the photometry of S190814bv. More specifically, we show
our \texttt{POSSIS} results alongside the EM upper limits of S190814bv in
three selected bands: ${\rm i}$ (top panel), ${\rm z}$ (middle panel),
and ${\rm K}$ (bottom panel). In all cases, the first column corresponds
to a face-on viewing angle (\ie independent of azimuthal angle), while
the second and third columns show the edge-on case with maximal (second
column) and minimal azimuthal (third column) angles, averaged as
described previously for
Fig.~\ref{fig:possis_photometric_luminosities_diff_filters}. The shaded
uncertainty regions around each light-curve reflect the uncertainty in
the distance. 

What is immediately clear, is that for all binaries the synthetic
light-curves lie systematically below the observational magnitude limits
obtained by the Dark Energy Camera (DECam) and VISTA-wide. This is true
even when considering the uncertainty in the distance from S190814bv,
which reflects differences of $\sim 0.3$-$0.5\,{\rm mag}$ at most. At the
same time, some differences emerge in presented bands. In particular, in
the ${\rm z}$-band and for the edge-on orientation -- chosen around the
azimuth that maximizes the cumulative luminosity -- the difference
between the upper bound at $t\approx 3.5\,\rm{d}$ and the light-curves of
the \texttt{Q4.chi0.8} and \texttt{Q5.chi0.8} binaries is the smallest
among all bands and orientations inspected here.  Interestingly, the
comparisons in the ${\rm K}$-band show, on average, the smallest and
systematic difference between the synthetic light-curves and the
observational limits. Furthermore, both weak- (\eg \texttt{Q7.chi0.8})
and strong-disruption mergers (\eg \texttt{Q4.chi0.8}) feature an EM
emission that is $\sim 1\,\rm{mag}$ smaller than the non-detection
thresholds of EM counterpart to S190814bv at $t\approx 1\,\rm{d}$, but
only the face-on view for strongly-disruptive binaries maintains this
closeness at late times $t\approx 10\,\rm{d}$. By contrast, the edge-on
views are systematically fainter by at least one magnitude initially, and
two magnitudes at later epochs.

\section{Conclusions}
\label{sec:conclusions}

In this paper, which is the third in a series of papers analysing the
properties of BHNS merger simulations~\cite{Topolski2024b,
  Topolski2024c}, we have reported a systematic investigation of the
dynamical mass ejection focusing on how the properties of the ejecta
depend on the properties of the binary, namely, the BH spin $\chi_{_{\rm
    BH}}$ and the mass-ratio $Q > 1$, for the realistic and temperature
dependent DD2 EOS. The overarching goal of this study is of establishing,
whenever possible, clear correlations between the geometric, kinematic
and thermodynamic features of dynamical ejecta -- and hence the
$r$-process nucleosynthetic yields and the KN signals -- with the
underlying binary parameters.

Overall, our analysis has highlighted a clear and systematic dependence
of all ejecta-related observables on $Q$ and $\chi_{{\rm BH}}$. Indeed,
according to the tidal-disruption strength reported in
Tab.~\ref{tab:ID_properties}, it is possible to reveal monotonic trends
in the bulk-ejecta properties as reported in
Fig.~\ref{fig:ye_s_vel_histogram} and summarized in
Tab.~\ref{tab:sim_diag}. More specifically, decreasing the mass ratio or
increasing the BH spin leads to larger dynamical ejecta masses, as well
as broader $Y_{e}$ and $s$ distributions; at the same time, large mass
asymmetries feature higher asymptotic ejecta velocities should the
disruption channel not be suppressed. In turn, these trends propagate to
the angular structure of the outflows, where our spherical-surface
extractions reveal a predominantly equatorial, crescent-like geometry
whose angular width measured by looking at the $90\%$ containment regions
shrinks with growing ejecta mass, indicating a high degree of
(directional) ejecta collimation. At the same time, weaker disruption
events exhibit broader, contiguous azimuthal coverage. Interestingly but
not surprisingly, these differences in the ejecta properties are
subsequently inherited by the $r$-process nucleosynthetic yields. In
particular, variations in the entropy and velocity of the ejecta directly
modulate the neutron-to-seed ratio and thereby the relative production of
heavy nuclei. Implementing in a controlled manner cutoffs in entropy and
velocity distributions has allowed us to further isolate the causes
behind these variations. In this way, we are able to confirm that the
observed pattern of second- and third-peak abundances and the variations
between models are both deeply rooted in the ejecta thermodynamics.

Complementary correlations between the ejecta properties and the
astrophysical observables have also been established by computing
synthetic kilonova signals. In this case, we have found that models with
larger dynamical ejecta produce systematically brighter and longer-lived
light-curves, whose viewing-angle dependence reflects the latitudinal
localisation of the outflow. Although none of our simulations was
tailored to the GW-inferred parameters of the AT2017gfo and S190814bv
events, the comparisons with the bolometric and photometric luminosities
obtained in these two cases have revealed that luminosity and temporal
evolution range covered by our models naturally spans the regimes
relevant for current BHNS kilonova searches. Furthermore, while all of
the models computed from our simulations show luminosities that are
weaker than those measured for S190814bv, they are also consistent with
all of its detection thresholds in the optical and NIR wavelengths.

In summary, the work presented here complements and concludes the
exploration of a comprehensive space of BHNS binary parameters in terms
of the mass ratio and BH spin so as to construct a complete description
of the dynamical processes accompanying a BHNS binary merger. While our
focus here has been on the consequences that the merger has in terms of
EM observables and nucleosynthetic yields and has been aimed at a
systematic study of these binaries, it has also relied on two
approximations. The first one is that it has neglected the irradiation of
the ejected material by a neutrino flux. 
Recent neutrino-transport calculations indicate that this can protonize a
non-negligible fraction of the dynamical ejecta and thus affect its
detailed $Y_e$ distribution, particularly with consequences for the
lighter $r$-process nuclei. Our conclusions concerning the very
neutron-rich component and the production of the heavier $r$-process
nuclei are expected to be more robust, while a quantitative assessment of
the full abundance pattern requires simulations including neutrino
transport.
The second
approximation made here is represented by the duration of the
simulations, which has had us focus exclusively on the $r$-process
nucleosynthesis yields and KN from the purely dynamical ejecta
component. In this case, because in BHNS systems the secular ejecta may
be expected to constitute a significant part of the unbound material,
changes to our conclusions and modelling the complete EM counterpart can
only be made with simulations performed on much longer
timescales. Notwithstanding these considerations, both of these
approximations will be lifted in a forthcoming investigation focusing on
long-term post-merger evolutions, jet-launching, and a potential sGRB
signal~\cite{Topolski2026a}.

\begin{acknowledgments}
  We are grateful to J\"urgen Schaffner-Bielich and Stephan Rosswog for
  discussions on the $r$-process nucleosynthesis and to Giulia
  Stratta for discussions on the EM upper limits and NSBH
	counterpart candidates. Additionally, we would like to thank the 
	anonymous referee for their comments that improved the clarity of this manuscript.
  K.T. acknowledges support by the Deutsche Forschungsgemeinschaft (DFG,
  German Research Foundation) under Germany's Excellence Strategy --
  EXC 2121 ``Quantum Universe'' -- 390833306.
  Support comes from the ERC Advanced Grant ``JETSET:
  Launching, propagation and emission of relativistic jets from binary
  mergers and across mass scales'' (Grant No. 884631). S.T. gratefully
  acknowledges support from NASA award ATP-80NSSC22K1898.
  P.S. acknowledges the support of the State of Hesse within the Research
  Cluster ELEMENTS (Project ID 500/10.006) and the European Union's
  Horizon 2020 Programme under the AHEAD2020 project (grant agreement n.
  871158). M.B. acknowledges the Department of Physics and Earth Science
  of the University of Ferrara for the financial support through the FIRD
  2024 and FIRD 2025 grants. L.R. acknowledges the Walter Greiner
  Gesellschaft zur F\"orderung der physikalischen Grundlagenforschung
  e.V. through the Carl W. Fueck Laureatus Chair. The simulations were
  performed on HPE Apollo HAWK at the High Performance Computing Center
  Stuttgart (HLRS) under the grants BNSMIC and BBHDISKS. The Calea
  cluster at ITP Frankfurt was used to post-process the NR simulation
  data and carry out \texttt{POSSIS} and \texttt{SkyNet} computations.
\end{acknowledgments}

\bibliographystyle{apsrev4-2}
\bibliography{./aeireferences}

\appendix 
\section{How the yields depend on the ejecta properties}
\label{subsec:rprocess_nucleosynthesis_entropy_vel_dep}

Relating the distribution of the nucleosynthetic yields to the properties
of the ejected material is far from simple. This is because normally a
complex mapping is made from the latter to the former, with complex
nuclear-reaction codes performing the mapping. Hence, the inverse
mapping, \ie how the yields actually depend on the ejecta properties, is
not always clear. To counter this, in the following we discuss how
artificial changes in the ejecta properties impact the final distribution
of the abundances shown in Fig.~\ref{fig:rprocess_abundances}.

In the specific context of our simulations, where the dynamical ejecta
all have rather uniform $Y_e$ distributions with no major differences
present among the various binaries (see
Fig.~\ref{fig:ye_s_vel_histogram}), the variance of the final abundance
results must necessarily and non-trivially arise from the combined
distribution of density, the ejecta velocity (together dictating the
expansion timescale), as well as the specific entropy. Indeed, the
spread of our results in Fig.~\ref{fig:rprocess_abundances} can be
explained in a relatively straightforward manner by manipulating the
underlying distribution of dynamical ejecta. To that end, we select the
\texttt{Q4.chi0.8} simulation as the one with the highest specific
entropy measures (see the third row, columns $3$--$5$ for $\langle s
\rangle$, $s_{75}$, $s_{95}$ in Tab.~\ref{tab:dyn_ejecta_properties}) as
well as the prominent high-entropy tail in the distribution of the
material~\cf~Fig.~\ref{fig:ye_s_vel_histogram}. Such a binary also
features a broad distribution of velocity of the dynamical ejecta
material (see $\langle v\rangle$, $v_{75}$, and $v_{95}$ in
Tab.~\ref{tab:dyn_ejecta_properties}, as well as the distribution in
Fig.~\ref{fig:ye_s_vel_histogram}).

More specifically, the following changes are applied to the raw dynamical
ejecta data collected on the surface of a detector:
\begin{itemize}
\item for \texttt{Q4.chi0.8}, $17$ datasets are prepared by rejecting
  parcels of the unbound material where $s\geq s_{\rm max}$, for $s_{\rm
    max}\in [40,200]\,k_{\rm B}/\rm{baryon}$ at intervals of $\Delta s =
  10\,k_{\rm B}/\rm{baryon}$; increasing the upper bound $s_{\rm max}$
  brings back the high-entropy contribution present originally,
\item for \texttt{Q4.chi0.8}, $17$ datasets of modified data are
  prepared, which require that unbound fluid elements are discarded if
  their velocity is smaller than $v_{\rm min}$, with $v_{\rm
    min}\in[0.15,0.55] $ steadily growing at intervals of $\Delta v_{\rm
    min} = 0.025\,c$; increasing $v_{\rm min}$ implies that a
  progressively larger fraction of the total ejecta mass is composed of
  high-velocity material.
\end{itemize}
For each of the modified datasets, $\approx 6000$--$8000$ tracers are
generated on average, and the expansion timescales and densities are
corrected for NSE in the same manner as in
Sec.~\ref{sec:r-process_yields}, so that the results presented here are
directly comparable with those of Fig.~\ref{fig:rprocess_abundances}. In
particular, the first $r$-process peak remains correctly absent in all of
the modified datasets, as expected from the intrinsic neutron richness
(\ie the uniformly low $Y_e$) of our dynamical ejecta.

As a result of these modifications, and following the procedure described
in Sec.~\ref{sec:r-process_yields} about the calculation of the
nucleosynthetic yields, we obtained $34$ artificially altered ejecta
datasets that can be used to unambiguously probe the influence of both
the inclusion of high-entropy material (case $1$) and the growing weight
(ratio) of high-velocity material in the total amount of the ejecta (case
$2$), or equivalently, the lack of low-velocity contributions. Even
though the altered ejecta datasets are clearly less massive than their
original counterparts due to rejecting some parts of the material via the
selection procedure, there still appears to be enough material to probe
the influence of $s$ and $v$ quantitatively.

\begin{figure*}
  \centering
  \includegraphics[width=1.0\textwidth]{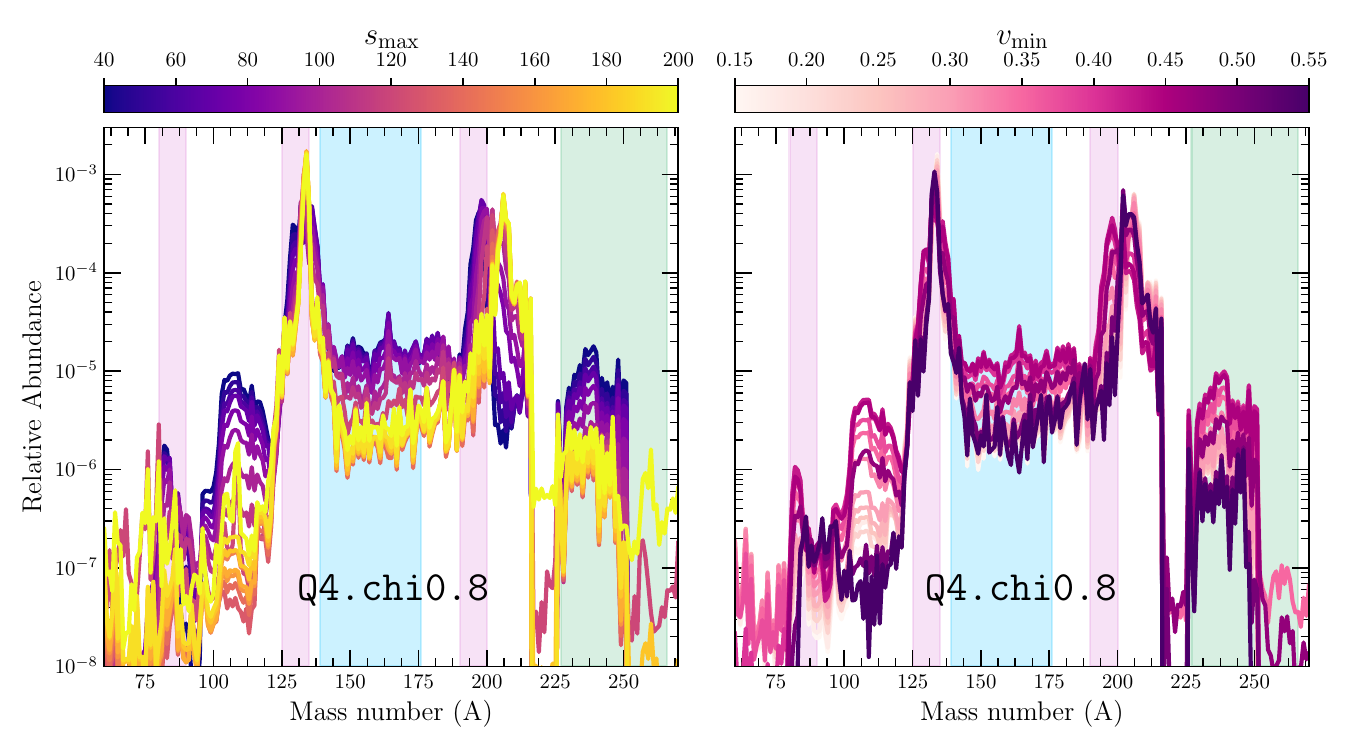}
  \caption{The same as in Fig.~\ref{fig:rprocess_abundances} with
    dynamical ejecta data of \texttt{Q4.chi0.8}, but where we
    artificially vary the upper bound for the specific entropy $s_{\rm
      max}$ (left panel) and the minimum velocity $v_{\rm min}$ (right
    panel) so as to highlight the impact of these cut-offs on the final
    distributions.}
  \label{fig:rprocess_abundances_cutoff}
\end{figure*}

The results of this broad investigation of the impact of the ejecta
properties on the nucleosynthetic yields is offered in
Fig.~\ref{fig:rprocess_abundances_cutoff}, whose left panel shows that
-- in our dynamical-ejecta-specific data -- the second and third
$r$-process peaks are remarkably robust against the entropy content of
the material, with only a modest change in the amplitude and a slight
shift of the third peak as $s_{\rm max}$ is varied. By contrast, the
abundances of the light $r$-process nuclei with $100 \lesssim A \lesssim
120$, of the inter-peak (trough) regions, and of the actinides are all
very sensitive to the entropy cut. More specifically, these regions are
most prominent when only the lowest-entropy material is retained: when
increasing $s_{\rm max}$ from $40\,k_{\mathrm{B}}/\mathrm{baryon}$ to
$200\,k_{\mathrm{B}}/\mathrm{baryon}$, and hence progressively
re-including the high-entropy tail of the distribution, the relative
abundances at $A \sim 110$ are suppressed by up to two orders of
magnitude, while the material between the second and third peaks and in
the actinide region decreases by roughly an order of magnitude. This
behavior reflects the fact that the low-entropy, slowly expanding
component of the ejecta is the one hosting the mildly incomplete
$r$-process flows that populate the regions between the main peaks and
sustain the production of the heaviest nuclei; once the high-entropy
tail is admitted, its strongly peaked contribution -- concentrated at
the second and third peaks -- dominates the (mass-weighted and
normalised) pattern and dilutes the relative abundances everywhere
else. 

Another summarising view of the impact of the ejecta properties on the
abundances is collected in the right panel of
Fig.~\ref{fig:rprocess_abundances_cutoff}, which illustrates the effects
of decreasing the fraction of ``slow'' ejecta in the total mass by
imposing a higher artificial lower velocity bound $v_{\rm min}$. Also in
this case, the second and third $r$-process peaks remain broadly
preserved for all $v_{\rm min}$ values, changing in amplitude by less
than a factor of a few and exhibiting only small shifts in position. The
most systematic trend is instead found for the light $r$-process nuclei
at $100 \lesssim A \lesssim 115$, whose relative abundances grow steadily
-- by up to an order of magnitude -- as $v_{\rm min}$ is increased to $v_{\rm min} \approx 0.45\,c$,
only to rapidly drop by $2$ orders of magnitude once this threshold is crossed, likely
due to sampling effects discussed below. 
This behavior is readily explained by the correlations in the ejecta: the
fastest material is also the one with the highest specific entropy
(\cf~Fig.~\ref{fig:Q7.chi0.8_slice_2d_xy}), so that removing the slow
component progressively shifts the composition toward high-entropy
conditions with reduced neutron exposure and favors the synthesis
of lighter $r$-process isotopes. Conversely, for
the most aggressive cuts, $v_{\rm min} \gtrsim 0.5\,c$, the abundances
between the peaks and in the actinide region are strongly suppressed,
dropping by roughly an order of magnitude with respect to the mildest
cuts; interestingly, this suggests that not enough high-velocity material
with $v \gtrsim 0.5\,c$ is present upon truncation to faithfully assess
the $r$-process yields in this regime.

We should also remark that this behavior is in part due to a selection
effect. More specifically, even if the number of tracers still samples
the modified ejecta distribution adequately, the associated mass weights
depend on correlations between thermodynamic and kinematic
quantities. Discarding ejecta with $v < v_{\rm min}$ can inadvertently
remove an associated region of phase-space in entropy (and $Y_e$) that is
strongly coupled to the slow tail of the velocity distribution. Since
this low-velocity matter is often neutron-rich and entropy-poor, its
removal disproportionately eliminates conditions favorable for the
heaviest $r$-process production.

\end{document}